\documentclass[aps,twocolumn,pra,amsmath,amssymb,superscriptaddress,floats]{revtex4}

\usepackage{amsmath}	
\usepackage{gensymb}
\usepackage{hyperref}
\hypersetup{colorlinks=true}
\usepackage{upgreek}
\usepackage{graphics}
\usepackage{hyperref}
\usepackage{epsfig}
\usepackage{color}
\usepackage{bm}
\usepackage{float}
\usepackage{ulem} 
\usepackage{dsfont}                 
\usepackage{wasysym}

\usepackage{blindtext}

\usepackage{accents}

\usepackage{graphicx}

\graphicspath{{./}{./plots/}}

\usepackage{tikz}
\usetikzlibrary{calc}

\usepackage{url}
\usepackage{multirow}

\newcommand{\upperRomannumeral}[1]{\uppercase\expandafter{\romannumeral#1}}

\begin{document}

\title{Unveiling the phase diagram of a bond-alternating spin-$\frac12$ $K$-$\Gamma$ chain}

\author{Qiang Luo}
\affiliation{Department of Physics, University of Toronto, Toronto, Ontario M5S 1A7, Canada}
\author{Jize Zhao}
\email[]{zhaojz@lzu.edu.cn}
\affiliation{School of Physical Science and Technology $\&$ Key Laboratory for Magnetism and
Magnetic Materials of the MoE, Lanzhou University, Lanzhou 730000, China}
\affiliation{Lanzhou Center for Theoretical Physics, Lanzhou University, Lanzhou 730000, China}
\author{Xiaoqun Wang}
\email[]{xiaoqunwang@sjtu.edu.cn}
\affiliation{Key Laboratory of Artificial Structures and Quantum Control (Ministry of Education), School of Physics and Astronomy, Tsung-Dao Lee Institute, Shanghai Jiao Tong University, Shanghai 200240, China}
\affiliation{Beijing Computational Science Research Center, Beijing 100084, China}
\author{Hae-Young Kee}
\email[]{hykee@physics.utoronto.ca}
\affiliation{Department of Physics, University of Toronto, Toronto, Ontario M5S 1A7, Canada}
\affiliation{Canadian Institute for Advanced Research, Toronto, Ontario, M5G 1Z8, Canada}

\date{\today}

\begin{abstract}
  The key to unraveling intriguing phenomena observed in various Kitaev materials lies in understanding the interplay
  of Kitaev ($K$) interaction and a symmetric off-diagonal $\Gamma$ interaction.
  To provide insight into the challenging problems, we study the quantum phase diagram of a bond-alternating spin-$1/2$ $g_x$-$g_y$ $K$-$\Gamma$ chain
  by density-matrix renormalization group method where $g_x$ and $g_y$ are the bond strengths of the odd and even bonds, respectively.
  The phase diagram is dominated by even-Haldane ($g_x > g_y$) and odd-Haldane ($g_x < g_y$) phases
  where the former is topologically trivial while the latter is a symmetry-protected topological phase.
  Near the antiferromagnetic Kitaev limit, there are two gapped $A_x$ and $A_y$ phases characterized by distinct nonlocal string correlators.
  In contrast, the isotropic ferromagnetic (FM) Kitaev point serves as a multicritical point where two topological phase transitions meet.
  The remaining part of the phase diagram contains three symmetry-breaking magnetic phases.
  One is a six-fold degenerate FM$_{U_6}$ phase where all the spins are parallel to one of
  the $\pm \hat{x}$, $\pm \hat{y}$, and $\pm \hat{z}$ axes in a six-site spin rotated basis,
  while the other two have more complex spin structures with all the three spin components being finite.
  Existence of a rank-2 spin-nematic ordering in the latter is also discussed.
\end{abstract}

\pacs{}

\maketitle

\section{Introduction}

The enigmatic quantum spin liquid~(QSL) has drawn a lot of attention
ever since the seminal work of Anderson in 1973 \cite{Anderson1973}.
In 2006, Kitaev proposed an exactly solvable spin-$1/2$ model on the honeycomb lattice
and demonstrated that its ground state is an exotic QSL
with emergent Majorana fermion excitations \cite{Kitaev2006}.
The past decade has witnessed a surge of interest in realization of the Kitaev honeycomb model
on real materials with $4d$ or $5d$ magnetic ions, which includes iridates and $\alpha$-RuCl$_3$
(see Refs.~\cite{Jackeli2009,RauLeeKee2016,TakagiTJ2019} and referees therein).
However, because of the inevitable non-Kitaev interactions,
e.g., the Heisenberg interaction and a symmetric off-diagonal exchange $\Gamma$-interaction \cite{RanLeeKeePRL2014},
these materials are shown to display magnetic orders at lowest temperatures
\cite{LiuBYetal2011,ChaloupkaJH2013,PlumbCSetal2014,JohnsonWHetal2015}.
Nevertheless, it is believed that the effective $K$-$\Gamma$ model is the dominant ingredient
to describe $\alpha$-RuCl$_3$ \cite{WangDYLi2017}.

From a theoretical point of view,
although the quantum phase diagram of the $K$-$\Gamma$ model on a honeycomb lattice is elusive,
several magnetically ordered phases and distinct QSLs are demonstrated to exist
\cite{RanLeeKeePRL2014,CatunYWetal2018,GohlkeWYetal2018,WangBL2019,LuoZhaoKeeWang2019,YamadaSS2020},
indicating the strong quantum fluctuation enhanced by competing interactions.
Given the notorious difficulty in two dimension,
it is beneficial and constructive to reduce the dimensionality where many full-fledged analytical and numerical methods
capable of addressing problems in one-dimensional~(1D) quantum spin chains are available.
Recently, the phase diagram of the isotropic $K$-$\Gamma$ chain has been studied
by the density-matrix renormalization group~(DMRG) method
and the non-Abelian bosonization technique \cite{YangKG2020,YangSN2021}.
It is shown that about 2/3 of the phase diagram is occupied by a gapless Luttinger liquid (LL).
The ferromagnetic~(FM) Kitaev limit is merely a transition point,
while a critical segment near the antiferromagnetic~(AFM) Kitaev limit is identified.
Two symmetry-breaking phases termed the FM$_{U_6}$ phase and the $M_2$ phase
(see Fig.~\ref{FIG-GSPD} for the nomenclature of the magnetically ordered phases) are also reported.
Later on, it is found that FM and AFM Heisenberg interactions could open up a wide region
of the FM$_{U_6}$ phase and the LL, respectively \cite{YangJKG2020}.
However, how to enlarge the territory of the puzzling $M_2$ phase is still unclear.
Aligning with this effort, a two-leg $K$-$\Gamma$ ladder under a [111] magnetic field is also studied,
revealing a rich phase diagram with several emergent phases \cite{SorenseCGK2021}.

Aside from the exotic phases and quantum criticality,
quantum spin chains also provide an excellent platform for theoretical studies of various quantum phase transitions~(QPTs) \cite{Sachdev2011},
of particular interest is the topological QPT that is beyond Landau's paradigm.
The topological QPT occurs between two different phases
without any explicit symmetry breaking \cite{KT1973,TsuiSG1982,HaldanePRL1983,WenRMP2017}.
The Haldane phase is such an example of symmetry-protected topological~(SPT) phase \cite{PollmannSPT2012},
which possesses a nonlocal string order parameter~(SOP) due to a hidden $\mathbb{Z}_2\times\mathbb{Z}_2$ symmetry breaking \cite{denNijsRom1989,KennedyTasaki1992},
Dating back to 1992, Hida originally pointed out that the bond-alternating spin-$1/2$ Heisenberg chain could host the Haldane phase
due to the imbalance of the neighboring coupling intensities,
leading to the formation of either total spin 0 or 1 out of the two spin-$1/2$ degrees of freedom \cite{Hida1992}.
Therefore, bond alternation is a practical route to legalize the validity of the SPT phase in spin-$1/2$ chains \cite{ChenGuWen2011}.
We also note that the anisotropic Kitaev spin chain hosts two disordered phases which undergo a direct transition at the isotropic point \cite{BrzezickiDO2007,YouTian2008}.
These observations motivate us to investigate the ground-state properties of the $K$-$\Gamma$ chain
by altering the bond strength of adjacent sites.

In this paper, we study the phases and QPTs of a bond-alternating $S$ = $1/2$ $K$-$\Gamma$ chain.
When $\Gamma = 0$, it is the Kitaev spin chain,
otherwise known as the exactly solvable 1D quantum compass model~(QCM) \cite{BrzezickiDO2007,YouTian2008}.
Beyond that it is nonintegrable except for some special points and lines when $|K|$ = $|\Gamma|$.
Therefore, we resort to the DMRG method \cite{White1992,Peschel1999,Schollwock2005}
to map out the quantum phase diagram.
The phase boundaries are determined by various quantities including the energy gap and entanglement entropy.
The central charge is calculated to distinguish the universality class of a continuous QPT.

The structure of the paper is as follows.
In Sec.~\ref{SEC:Model} we introduce the theoretical model under investigation,
analyse the symmetry properties, and present the phase diagram of interest.
Following this, we study two topological QPTs in Secs.~\ref{SEC:EHOH} and \ref{SEC:QCM}.
Section~\ref{SEC:SSB} presents the magnetic order parameters of symmetry-breaking phases.
In Sec.~\ref{SEC:FMKtv} we study the transitions between the left and right panels of the phase diagram.
We conclude with a summary in Sec.~\ref{SEC:Conclusion}.
Finally, a brief review of the diagonalization of QCM
and some other useful contents are presented in the Supplemental Material \cite{SuppMat}.

\section{Model and Method}\label{SEC:Model}
\begin{figure}[!ht]
\centering
\includegraphics[width=0.99\columnwidth, clip]{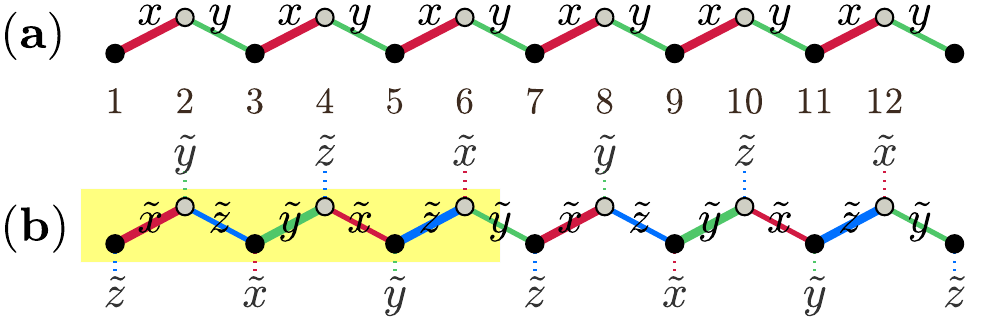}\\
\caption{(a) Sketch of the bond structure in the original form.
    Here, $x$~(red) and $y$~(green) stand for the $\gamma$-index
    and the width is proportional to the bond strength.
    (b) Pictorial bond structure of the Hamiltonian in the rotated basis.
    The overhanging bond at each site is determined by the remaining one along the chain.
    The shaded region represents the six-site unit cell.}
    \label{FIG-Bond}
\end{figure}

We consider a bond-alternating spin-$1/2$ $K$-$\Gamma$ chain with
\begin{align}\label{J1J2KG-Ham}
\mathcal{H} = \sum_{l=1}^{L/2} g_x\mathcal{H}_{2l-1,2l}^{(x)}(\theta) + g_y\mathcal{H}_{2l,2l+1}^{(y)}(\theta)
\end{align}
where $L$ is the chain length, $g_x$~($g_y$) is the odd~(even) bond strength, and
\begin{align}\label{KtvGam-Ham}
\mathcal{H}_{i,j}^{(\gamma)}(\theta) = K S_i^{\gamma}S_j^{\gamma} + \Gamma (S_i^{\alpha}S_j^{\beta}+S_i^{\beta}S_j^{\alpha}).
\end{align}
Here, $K$ and $\Gamma$ are the Kitaev interaction and the off-diagonal exchange interaction, respectively.
$\gamma$ could be either $x$ or $y$ and it specifies the spin direction associated with the referred bond, see Fig.~\ref{FIG-Bond}(a).
For each $\gamma$-bond, $\alpha$ and $\beta$ are the two remaining mutually exclusive spin directions.
In what follows we parametrize $K=\sin\theta$ and $\Gamma=\cos\theta$ with $\theta \in (-\pi, \pi]$.

\begin{figure}[!ht]
\centering
\includegraphics[width=0.90\columnwidth, clip]{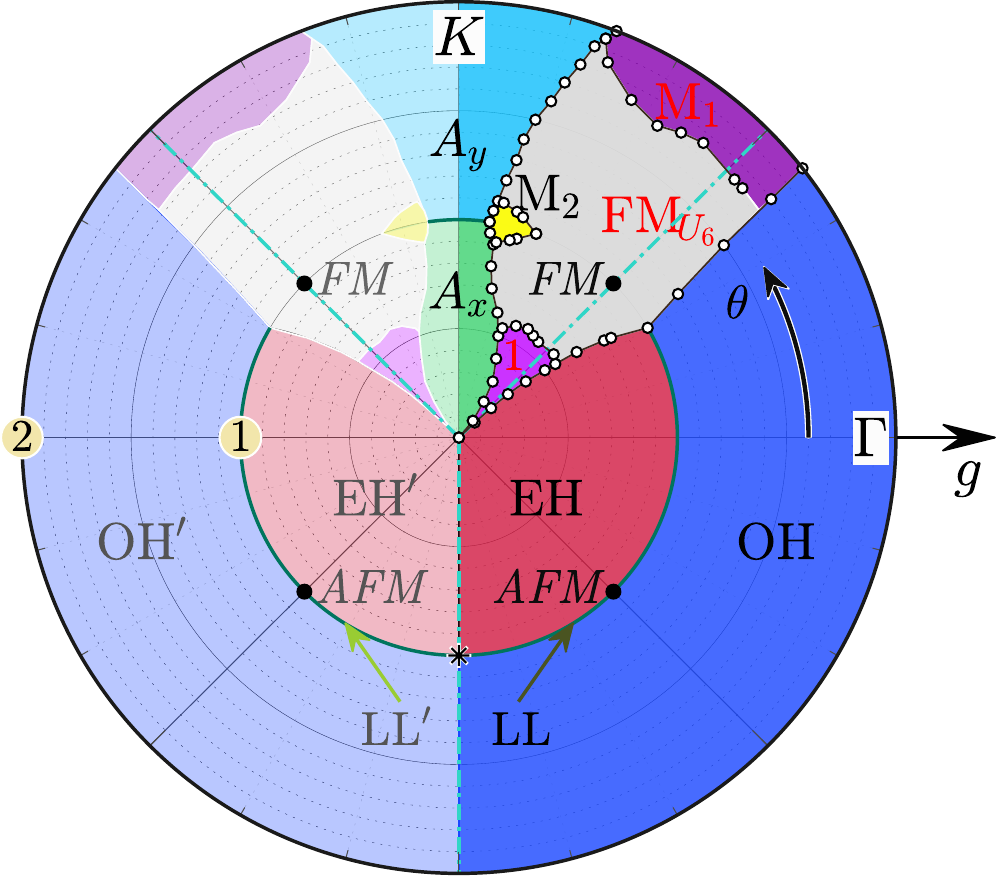}\\
\caption{Quantum phase diagram of the bond-alternating spin-$1/2$ $K$-$\Gamma$ chain with $K=\sin\theta$ and $\Gamma=\cos\theta$.
    The green thick line marked by $\textcircled{1}$ is the isotropic $K$-$\Gamma$ chain,
    and the black solid circles at $\theta = \pi/4$ and $-\pi/4$ represent the hidden $SU(2)$ FM and AFM Heisenberg chains, respectively.
    The asterisk ($\ast$) in the vertical line represents a multicritical point.
    There are seven distinct phases in the right panel which are the main focus of the paper.
    The EH--OH transition and $A_x$--$A_y$ transition are continuous with a central charge of $c = 1$ and $1/2$, respectively.
    The nature of the magnetically ordered states, the FM$_{U_6}$ phase, the $M_1$ phase, and the $M_2$ phase,
    are clarified in a six-site $U_6$ rotated basis.
    See the main text for details.}
    \label{FIG-GSPD}
\end{figure}

Before carrying out numerical calculation,
let us analyze the symmetries in the parameter space
which will reduce the computational cost.
Akin to the transverse field Ising model \cite{Pfeuty1970},
the model \eqref{J1J2KG-Ham} owns a duality relation which could be seen by applying the spin rotation transformation
$(S_i^x, S_i^y, S_i^z) \to (-S_i^y, -S_i^x, -S_i^z)$.
This implies that each eigenvalue $E$ of $H$ satisfies the relation
\begin{equation}\label{SelfDual}
E(g) = gE(1/g)
\end{equation}
where $g \equiv g_y/g_x$ is the relative bond strength.
On the other hand, by virtue of a global spin rotation around the $z$-axis by $\pi$,
$(S_i^x, S_i^y, S_i^z) \to (S_i^y, -S_i^x, S_i^z)$,
the Kitaev interaction remains uninfluenced whereas the sign of $\Gamma$-interaction is altered \cite{YangKG2020}.
We thus instantly find that
\begin{equation}\label{GammaSym}
E(K, \Gamma) = E(K, -\Gamma),
\end{equation}
or equivalently, $\theta \mapsto \pi-\theta$.
These relations in Eq.~\eqref{SelfDual} and Eq.~\eqref{GammaSym}
allow us to focus on the phase diagram primarily in the reduced parameter range
$\theta \in [-\pi/2, \pi/2]$ and $g = g_y/g_x \in [0, 1]$
and then map out the whole phase diagram shown in Fig.~\ref{FIG-GSPD}.

Using a site-ordering cross decimation rotation with a periodicity of six sites,
all the cross terms of $S_i^{\alpha}S_j^{\beta}$ with $\alpha \neq \beta$ in Eq.~\eqref{KtvGam-Ham} will vanish \cite{YangKG2020}.
This $U_6$ transformation is given by
\begin{eqnarray}\label{EQ:SOCD}
\text{sublattice $1$}: & (x,y,z) & \rightarrow (\tilde{x},\tilde{y},\tilde{z}),      \nonumber\\
\text{sublattice $2$}: & (x,y,z) & \rightarrow (-\tilde{x},-\tilde{z},-\tilde{y}),   \nonumber\\
\text{sublattice $3$}: & (x,y,z) & \rightarrow (\tilde{y},\tilde{z},\tilde{x}),      \nonumber\\
\text{sublattice $4$}: & (x,y,z) & \rightarrow (-\tilde{y},-\tilde{x},-\tilde{z}),   \nonumber\\
\text{sublattice $5$}: & (x,y,z) & \rightarrow (\tilde{z},\tilde{x},\tilde{y}),      \nonumber\\
\text{sublattice $6$}: & (x,y,z) & \rightarrow (-\tilde{z},-\tilde{y},-\tilde{x}),
\end{eqnarray}
where $\gamma \big(= x (\tilde{x}), y (\tilde{y}), z (\tilde{z})\big)$ denotes the spin component of $S^{\gamma}$ ($\tilde{S}^{\gamma}$).
Under this transformation the original Hamitonian acquires the following form \cite{YangKG2020}
\begin{align}\label{U1XYZ-Ham}
\tilde{\mathcal{H}}_{i,j}^{(\gamma)}(\theta) = -K \tilde{S}_i^{\gamma} \tilde{S}_j^{\gamma}
- \Gamma (\tilde{S}_i^{\alpha} \tilde{S}_j^{\alpha}+ \tilde{S}_i^{\beta} \tilde{S}_j^{\beta})
\end{align}
in which the bonds $\gamma$ = $\tilde{x}$ (red), $\tilde{z}$ (blue), and $\tilde{y}$ (green) circularly, as depicted in Fig.~\ref{FIG-Bond}(b).
$\tilde{\mathbf{S}} = (\tilde{S}_i^{x}, \tilde{S}_i^{y}, \tilde{S}_i^{z})$ is the spin operator in the rotated basis.
Such a $U_6$ transformation does not alter the energy spectra (i.e., energy and its degeneracy)
but simplifies the spin-spin correlation functions.
Therefore, we will preferentially focus on the rotated Hamiltonian in Eq.~\eqref{U1XYZ-Ham} unless stated explicitly otherwise.
The exceptions are Secs.~\ref{SEC:QCM} and \ref{SEC:FMKtv}
where it is convenient to calculate the correlation functions in the original basis.
In addition, combining Eq.~\eqref{GammaSym},
it is apparently that Eq.~\eqref{U1XYZ-Ham} has a $SU(2)$ symmetric structure
when $|K| = |\Gamma|$.
Specifically, In the range $\theta \in [-\pi/2, \pi/2]$,
the point $\theta = -\pi/4$ and $\pi/4$ corresponds to an
AFM and FM Heisenberg chain, respectively.

The numerical calculations are performed by the DMRG method
\cite{White1992,Peschel1999,Schollwock2005},
which is a powerful technique for 1D many-body problems.
Periodic boundary condition~(PBC) is preferred to weaken the finite-size effect
and open boundary condition~(OBC) is also adopted occasionally for comparison.
We keep up to 2000 states so as to ensure a typical truncated error of $\sim\!10^{-7}$ or less.
The chain length $L$ is strictly considered to be the multiple of 6,
consistent with the structure of unit cell and the $U_6$ transformation.

The resultant phase diagram is shown in Fig.~\ref{FIG-GSPD},
which has a salient feature of mirror (left-right) symmetry.
Focusing on the right half circle, it harbours \textit{seven} distinguishing phases.
Four of them, i.e., the even-Haldane~(EH) and odd-Haldane~(OH) phases
and the $A_x$ and $A_y$ phases \cite{Kitaev2006},
are disordered and could be characterized by nonlocal SOPs of different kinds.
The rest are three magnetically ordered phases named FM$_{U_6}$ phase and $M_1$ and $M_2$ phases.
The FM$_{U_6}$ phase is collinear in the rotated basis and exhibits six-fold degeneracy.
The $M_1$ and $M_2$ phases show more complex spin patterns where all their three spin components are finite.
For the $M_1$ phases, it is stabilized at the region where $g \lesssim 1/\sqrt3$ or $g \gtrsim \sqrt3$
and \textit{one} of the spin components is dominantly the biggest.
For the $M_2$ phases, it locates around the very isotropic line of $g \simeq 1$ where $K/\Gamma > 1$
and \textit{two} of its spin components are almost the same and is larger than the third.

\section{EH-OH topological QPT}\label{SEC:EHOH}

Straightforwardly, when $\theta = -\pi/4$ Eq.~\eqref{J1J2KG-Ham} and Eq.~\eqref{U1XYZ-Ham}
turn out to be a bond-alternating AFM Heisenberg chain \cite{Hida1992,Barnes1999,Johnston2000,WangLiCho2013,Haghshenas2014}.
It is well-established that there is a topological EH-OH transition at $g = 1$ with a central charge $c = 1$ \cite{WangLiCho2013}.
For either $g < 1$ or $g > 1$, the ground state could be characterized by a SOP
which is nonzero inside the phase but vanishes otherwise (see Sec. II in the Supplemental Material \cite{SuppMat}).
Specifically, the two phases could be distinguished by the even- and odd-SOPs which are defined as \cite{Hida1992}
\begin{equation}\label{EQ:SOPev}
\mathcal{O}_{e}^{\alpha} = \lim_{|j-i|\to\infty}\mathcal{O}^{\alpha}(2i,2j+1)
\end{equation}
and
\begin{equation}\label{EQ:SOPod}
\mathcal{O}_{o}^{\alpha} = \lim_{|j-i|\to\infty}\mathcal{O}^{\alpha}(2i-1,2j)
\end{equation}
where
\begin{equation}\label{EQ:SOPij}
\mathcal{O}^{\alpha}(p,q) = -4\left\langle \tilde{S}_p^{\alpha} \Big(\prod\limits_{p<r<q} e^{i\pi \tilde{S}_r^{\alpha}}\Big) \tilde{S}_q^{\alpha} \right\rangle.
\end{equation}
Here, $\alpha = x, y, z$.
The even-SOP $\mathcal{O}_{e}^{\alpha}$ is calculated from an even site ($2i$) to an odd site ($2j+1$)
while the odd-SOP $\mathcal{O}_{o}^{\alpha}$ is measured from an odd site ($2i-1$) to an even site ($2j$).
At the critical point $g = 1$, both SOPs are identical and decay as $\mathcal{O}_{e/o} \sim L^{-1/4}$ \cite{BortzSS2007}.
From a topological perspective, the EH phase is trivial while the OH is a SPT phase
which is isomorphic to the ground state of the spin-1 Heisenberg chain \cite{HaldanePRL1983}.
For the OH phase, its ground state is unique under PBC
but has a fourfold degeneracy under OBC because of two edge spin-$1/2$s.
In addition, the degeneracy of the lowest-lying entanglement spectrum
is twofold~(fourfold) under OBC~(PBC) \cite{Pollmann2010}.

\begin{figure}[!ht]
\centering
\includegraphics[width=0.95\columnwidth, clip]{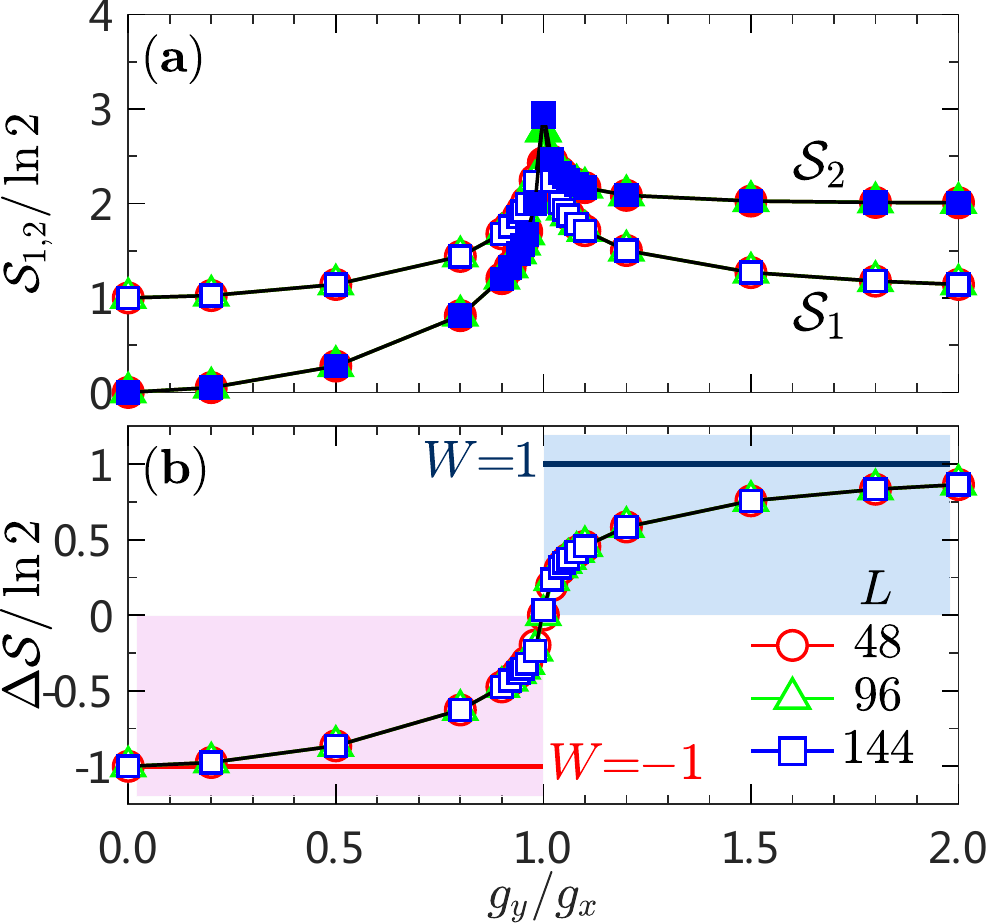}\\
\caption{(a) Entanglement entropy $\mathcal{S}_1$ (open symbols) and $\mathcal{S}_2$ (filled symbols) of different cuts
    for the $g_x$-$g_y$ $\Gamma$-chain with $\theta= 0.00\pi$.
    The chain length $L$ is 48~(red circle), 96~(green triangle), and 144~(blue square).
    (b) Bulk entanglement gap $\Delta\mathcal{S}$ in the same region as (a).
    $W = -1$ and 1 are the sign of $\Delta\mathcal{S}$ when $g < 1$ and $g > 1$, respectively.}
    \label{FIG-BulkDltVNE}
\end{figure}

As shown in Fig.~\ref{FIG-GSPD},
the EH and OH phases extend to a large region of the parameter space.
To demonstrate it, we focus on the line of $\theta = 0$, which is the $\Gamma$-chain limit.
We begin by studying a so-called \textit{bulk entanglement gap} $\Delta\mathcal{S}$ \cite{Tan2020},
which comes from the even-odd oscillation of the entanglement entropy
$\mathcal{S}_L(l) = -\mathrm{Tr}(\rho_l\ln\rho_l)$
where $\rho_l$ is the reduced density matrix of the subsystem
with a contiguous spatial segment $l$ \cite{VidalLRK2003}.
Depending on whether $l \gg 1$ is odd or even,
$\mathcal{S}_L(l)$ saturates to a constant value of $\mathcal{S}_1$ and $\mathcal{S}_2$, respectively.
The bulk entanglement gap is thus defined as $\Delta\mathcal{S} = \mathcal{S}_2 - \mathcal{S}_1$ \cite{Tan2020}.
Figure~\ref{FIG-BulkDltVNE}(a) shows $\mathcal{S}_1$~(open symbols) and $\mathcal{S}_2$~(fill symbols)
for the chain length $L$ = 48, 96, and 144.
For these lengths chosen, $\mathcal{S}_1$ corresponds to cut one strong valence bond consistently,
while $\mathcal{S}_2$ stands for cutting a strong valence bond
zero time or twice when $g < 1$ or $g > 1$, respectively.
This implies that the  bulk entanglement gap $\Delta\mathcal{S}/\ln2$ tends to be $-1$ or $1$ at the limit case
where $g\to0$ or $g\to\infty$, respectively~(see Fig.~\ref{FIG-BulkDltVNE}(b)).
Near the critical point, $\mathcal{S}_1 \simeq \mathcal{S}_2$ and thus $\Delta\mathcal{S} \simeq 0$.
Moreover, defining $\delta = \frac{g-1}{g+1}$,
the quantity scales as $\Delta\mathcal{S} \sim \ln2-(1-|\delta|)^2$ when away from criticality,
whereas $\Delta\mathcal{S} \sim -\delta\ln|\delta|$ when close to criticality \cite{Tan2020}.
Therefore, the value of $\Delta\mathcal{S}$ is bounded to $\pm\ln2$
and its sign $W = \textrm{sgn}\big(\Delta\mathcal{S}\big)$
could be used to characterize the corresponding topological sector.
The sign change~(e.g., from $W = -1$ to 1) is a signal for the topological QPT.

\begin{figure}[!ht]
\centering
\includegraphics[width=0.95\columnwidth, clip]{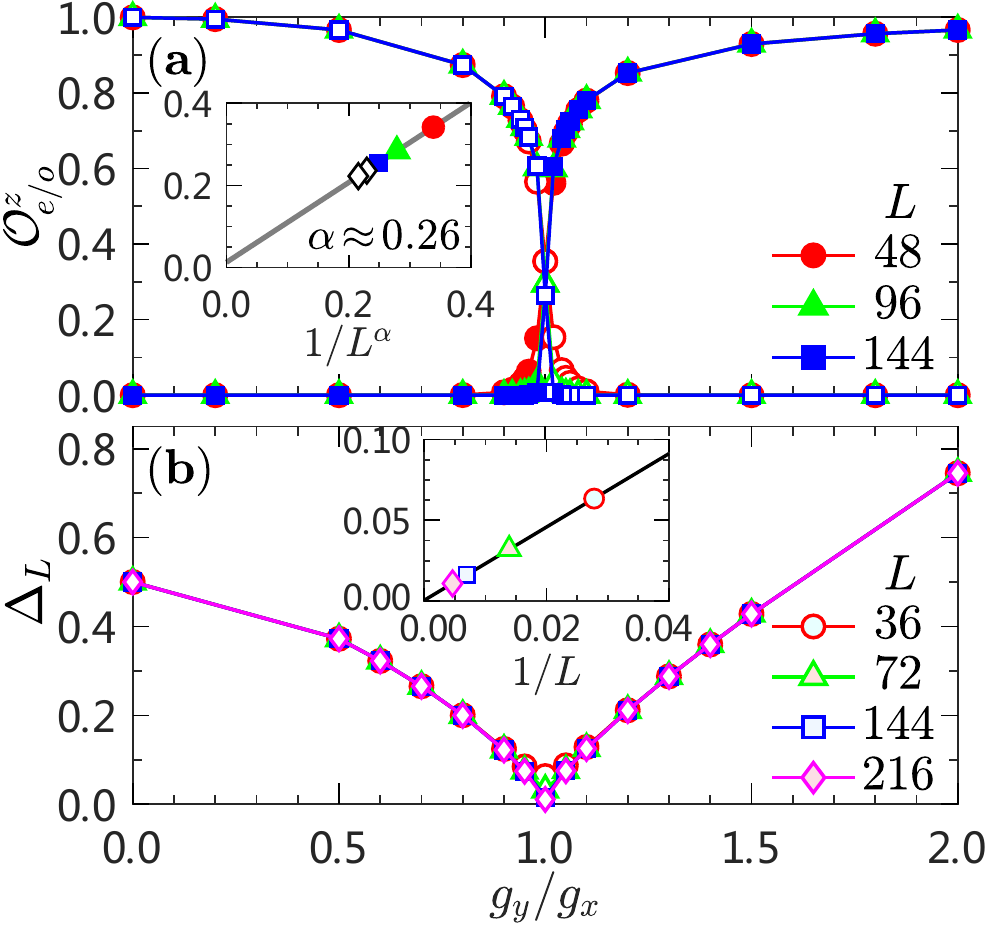}\\
\caption{(a) SOPs of the even type $\mathcal{O}_{e}^z$ (open symbols) and odd type $\mathcal{O}_{o}^z$ (filled symbols)
    for the $g_x$-$g_y$ $\Gamma$-chain with $\theta= 0.00\pi$.
    The inset shows the asymptotic decay of SOP $\mathcal{O}$ at $g_x = g_y$.
    (b) Energy gap $\Delta_L$ in the same region as (a).
    Inset shows the linear extrapolation of the energy gap at $g_x = g_y$.}\label{FIG-GamSOPGap}
\end{figure}

To further reveal the nature of phases at different topological sectors,
we measure the even-SOP $\mathcal{O}_{e}^{z}$ (see Eq.~\eqref{EQ:SOPev})
and odd-SOP $\mathcal{O}_{o}^{z}$ (see Eq.~\eqref{EQ:SOPod}).
It is clearly shown in Fig.~\ref{FIG-GamSOPGap}(a) that
$\mathcal{O}_{e}^{z}$~($\mathcal{O}_{o}^{z}$) is finite when $g < 1$~($g > 1$) and is vanishingly small otherwise.
The finite-size effects of $\mathcal{O}_{e/o}^{z}$ are very weak,
except for a narrow window that is close to the critical region.
As shown in the inset, both types of SOPs $\mathcal{O}_{e/o}^{z}$ decay algebraically as $L^{-\alpha}$
where the critical exponent $\alpha \approx 0.26$, which is fairly close to the value of $1/4$ at $\theta = -\pi/4$ \cite{BortzSS2007}.
For an infinite size system, SOPs $\mathcal{O}_{e/o}^{z}$ scale as $\delta^{1/6}$ \cite{Hida1992}.
As a result, the critical exponent is given as $\beta = 1/12$ because $\mathcal{O} \propto \delta^{2\beta}$.
Hence, this topological QPT belongs to the Gaussian universality class.
We also calculate the excitation gap $\Delta_T$,
which is defined as the energy difference between the first excited state and the ground state.
Figure~\ref{FIG-GamSOPGap}(b) shows that $\Delta_T$ is very robust when $g \neq 1$.
Near $g = 1$, it has a pronounced drop with size increased.
As shown in the inset, $\Delta_T$ is zero when $L\to\infty$,
showing that the ground state of the isotropic $\Gamma$-chain is critical.
To extract the central charge $c$, we calculate the von Neumann entanglement entropy $\mathcal{S}_L(L/2)$
for a series of chain length $L$ and the central charge is fitted by
$\mathcal{S}_L = \frac{c}{3}\ln(L/\pi) + c'$. 
Our best fitting suggests that $c \simeq 0.997(5)$ (not shown),
which is very close to 1 of the LL.

\begin{figure}[!ht]
\centering
\includegraphics[width=0.95\columnwidth, clip]{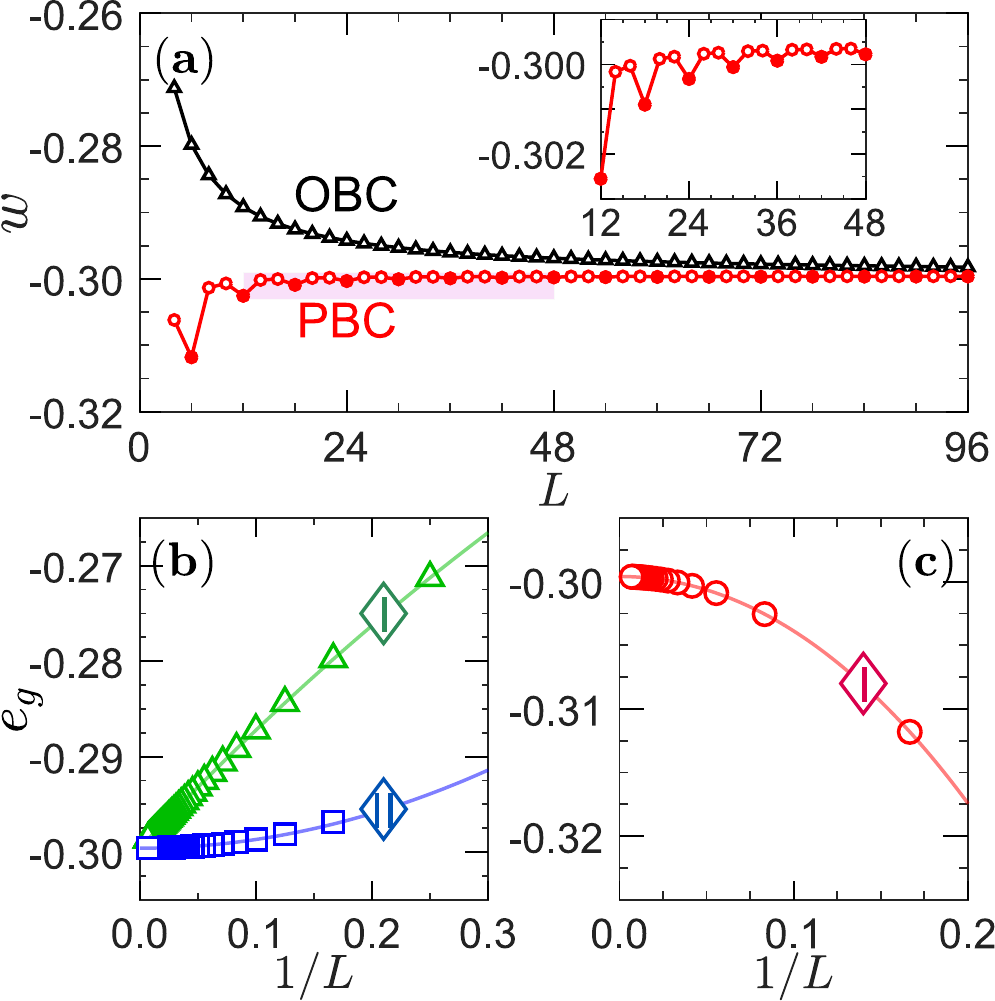}\\
\caption{(a) Behaviors of the energy density $w = E_g(L)/L$ for the isotropic $\Gamma$-chain under OBC~(black) and PBC~(red).
    The inset shows the six-site periodicity of $w$ under PBC.
    (b) and (c) show the estimate of $e_g$ for the isotropic $\Gamma$-chain under OBC and PBC, respectively.
    The roman numerals \upperRomannumeral{1} (green and red) and \upperRomannumeral{2} (blue)
    mark two different methods illustrated in the text.}\label{FIG-GamEg}
\end{figure}

Figure~\ref{FIG-GamEg}(a) displays the energy density $E_g/L$ of the isotropic $\Gamma$-chain
under OBC (black triangle) and PBC (red circle).
For the OBC case, $E_g/L$ decreases smoothly and saturates around $-0.30$ as $L$ increases.
In contrast, it is not monotonically increasing but exhibits an oscillation with six-site periodicity for the PBC (see inset).
As a comparison, we note that such an abnormal energy density behavior is absent in the isotropic Kitaev spin chain
(see Fig. 1 in the Supplemental Material \cite{SuppMat}).
This phenomenon in the $\Gamma$-chain is striking and may be related to the unusual energy behavior
of the $\Gamma$ model on the honeycomb lattice \cite{LuoZhaoKeeWang2019}.
In Ref.~[\onlinecite{LuoZhaoKeeWang2019}] the total energy $E_g$ is calculated on a series of honeycomb clusters
where OBC~(PBC) is utilized on the $L_x$~($L_y$) direction of cylinders.
For any cylinder with fixing $L_y$, the energy density $E_g/N$ ($N=L_xL_y$) varies linearly with $1/L_x$.
However, by increasing the circumference of the cylinders with $L_x/L_y = 2$,
the energy density $E_g/N$ is no longer monotonous and exhibits a skew sawtooth behavior.

To round off the calculation, we give an estimate of the ground-state energy per-site $e_g$ of the isotropic $\Gamma$-chain.
We note that our $\Gamma$-chain contains both $x$ and $y$ bonds~(see Eq.~\eqref{J1J2KG-Ham}),
and there is no analytical solution so far.
It is fundamentally different from a $z$-bond $\Gamma$-chain
which could be solved exactly via the Jordan-Wigner transformation \cite{YouGam2020}.
At the quantum critical point, the finite-size scaling of the ground-state energy $E_g(L)$ is known to be \cite{BloteCN1986,Affleck1986}
\begin{equation}\label{EQ:EgFSS}
E_g(L) = Le_g + \varepsilon_b - \frac{\Delta_b}{L} + \mathcal{O}(L^{-2}),
\end{equation}
where $e_g$ is the average bulk energy per-site,
$\varepsilon_b$ is the size-independent surface energy which vanishes in the case of PBC,
and $\Delta_b$ is the subleading correlation term.
It is found that $\Delta_b = \pi c/6$~($\pi c/24$) for PBC~(OBC) where $c$ is the central charge \cite{BloteCN1986,Affleck1986}.
By definition we have $e_g = \lim_{L\to\infty} e_L$
where $e_L$ is the energy per-site of the chain with length $L$.
For the energy obtained under the OBC,
there are two ways to extrapolate it to the thermodynamic limit;
one is $e_L^{\textrm{I}} = E_g(L)/L$ and the other is $e_L^{\textrm{II}} = \big(E_g(L)-E_g(L\!-\!2)\big)/2$.
It is easy to check that convergence speed of the latter is faster than the former.
As shown in Fig.~\ref{FIG-GamEg}(b), the quadratic fittings of the two give that
$e_g^{\textrm{I}} = -0.29959362$ and $e_g^{\textrm{II}} = -0.29959375$,
yielding an estimate for the ground-state energy per-site in the thermodynamic limit
of $e_g = -0.2995937(1)$ with seven significant digits.
Meanwhile, we also extrapolate the energy under PBC
by using the solid points in Fig.~\ref{FIG-GamEg}(a) where $L$ is a multiple of six~(see Fig.~\ref{FIG-GamEg}(c)).
Our result suggests that $e_g \approx -0.299594$,
which is fairly consistent with the high-precision value revealed by the calculation under OBC.

\section{Extended quantum compass model}\label{SEC:QCM}

In the absence of $\Gamma$-interaction,
Eq.~\eqref{J1J2KG-Ham} and Eq.~\eqref{KtvGam-Ham} are reduced to the 1D Kitaev spin chain,
which is also known as the 1D QCM in some other context \cite{BrzezickiDO2007,YouTian2008}.
The QCM could be solved exactly by Jordan-Wigner transformation
and its dispersion relation is almost the same as that of the transverse field Ising model.
There is a topological QPT between the gapped $A_x$ and $A_y$ phases at $g = 1$ \cite{Kitaev2006}.
However, because of intermediate symmetries,
the ground state of the QCM possesses a huge number of degeneracy $2^{N/2-1}$~($2^{N/2}$) under PBC~(OBC)
where $N$ is the total number of sites \cite{BrzezickiDO2007,NussinovBrink2015}.
Equivalently, the QCM could be rewritten as a Majorana fermion chain complemented by $N$ decoupled Majorana fermions.
Since each Majorana fermion has $\sqrt2$ degrees of freedom,
the redundant Majorana fermions thus contribute a ground-state degeneracy of $\mathcal{O}(2^{N/2})$ \cite{FengZX2007}.
These degenerate ground states are vulnerable and can be totally lifted by an infinitesimal transverse field \cite{SunChen2009}.
The entanglement \cite{Subrahmanyam2013,YouWangYi2018,VimalSubra2018,WuYou2019},
energy dynamics \cite{SteinBrenig2016}, and the dissipative behavior \cite{ShibataKat2019}
of the QCM have been studied over the years.

Using the spin duality transformation, the topological $A_x$ and $A_y$ phases, respectively,
are shown to possess nonlocal string correlators \cite{FengZX2007},
\begin{equation}\label{EQ:SOPx}
\mathcal{O}_K^x(2r) = \lim_{r\to\infty}\left\langle\prod_{k=1}^{2r} \sigma_k^x\right\rangle
\end{equation}
and
\begin{equation}\label{EQ:SOPy}
\mathcal{O}_K^y(2r) = \lim_{r\to\infty}\left\langle\prod_{k=2}^{2r+1} \sigma_k^y\right\rangle,
\end{equation}
where $\sigma_k^x$ and $\sigma_k^y$ are Pauli matrices, i.e., twice the spin-1/2 operators in the \textit{original} basis.
Here, the nonlocal SOPs are defined in the original basis embedded in Eq.~\eqref{KtvGam-Ham}.
Generalization of nonlocal SOPs to a two-leg Kitaev ladder has been discussed in a recent work \cite{Catuneanu2019}.
At the critical point $g = 1$, these Kitaev-type SOPs vanish in an algebraic behavior at long-distance limit $n\gg1$ \cite{Pfeuty1970},
\begin{equation}\label{EQ:KSOPAsy}
\mathcal{O}_K^{x/y}(n) =e^{1/4}2^{1/12}A^{-3} n^{-1/4}\Big(1-\frac{1}{64}n^{-2} + \cdots\Big),
\end{equation}
where $A \simeq 1.2824$.
For infinite-size case, they obey a scaling law and $\mathcal{O}_K^{x} \sim (1-g^2)^{1/4}$ when $g\to1^{-}$.

\begin{figure}[!ht]
\centering
\includegraphics[width=0.95\columnwidth, clip]{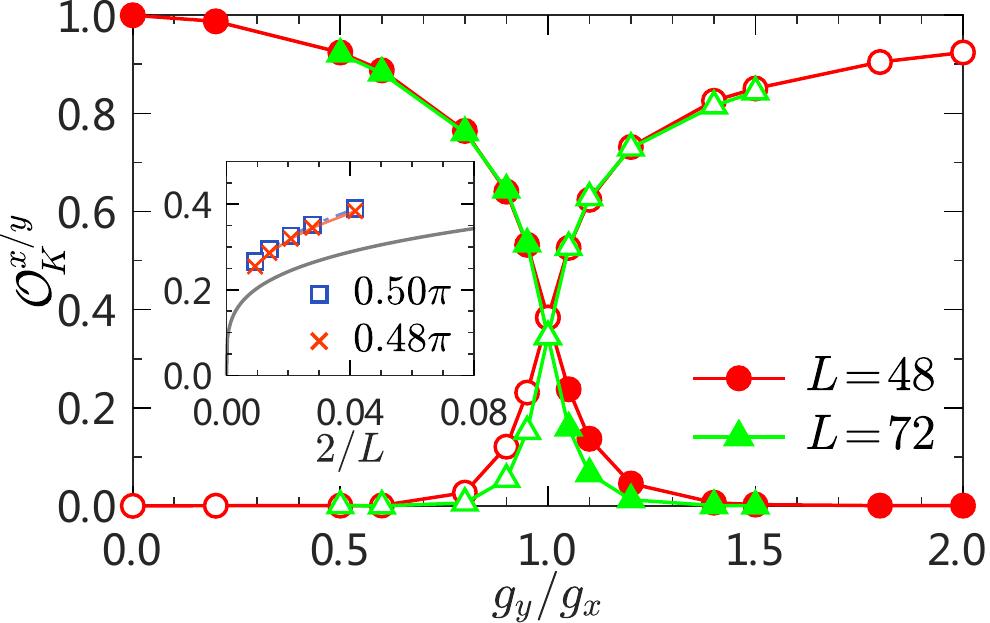}\\
\caption{Kitaev-type SOP $\mathcal{O}_K^{x/y}$ of the $g_x$-$g_y$ $K$-$\Gamma$ chain for $\theta=0.48\pi$
    with chain length $L = 48$~(red circle) and 72~(green triangle).
    The inset shows the behavior of $\mathcal{O}_K^{x}$ at the isotropic point $g = 1$.
    Values at $\theta = 0.50\pi$~(blue square) and $\theta = 0.48\pi$~(red cross) are shown
    for several chain length $L$ = 48, 72, 96, 144, and 216.
    The sold line is the correlation function defined in Eq.~\eqref{EQ:KSOPAsy}.}
    \label{FIG-Tht048}
\end{figure}

Hereafter we show numerically that,
the topological $A_x$ and $A_y$ phases are extended
when $\theta$ is slightly deviated from $\pi/2$ (AFM Kitaev point).
We demonstrate this by calculating the SOPs shown in Eq.~\eqref{EQ:SOPx} and Eq.~\eqref{EQ:SOPy} for $\theta = 0.48\pi$ with $2r = L/2$.
As presented in Fig.~\ref{FIG-Tht048},
the SOPs change smoothly and are very robust in each corresponding phase,
showing the validity of them in this region.
When $g = 1$, the two have the same value due to the self-dual relation and they decrease visibly as $L$ grows.
To measure how the SOPs vary at this point,
we calculate $\mathcal{O}_K^{x}$ for $L$ up to 216 sites and the results are shown in the inset.
The values at $\theta = 0.50\pi$ are also shown for comparison.
As revealed by Eq.~\eqref{EQ:KSOPAsy},
the leading term of $\mathcal{O}_K^{x}$ is $\sim1/L^{1/4}$,
so its decay ratio is not very rapid for modest chain length $L$ as shown by the solid line in the inset.
However, it is constructive to note that $\mathcal{O}_K^{x}$ at $\theta = 0.48\pi$ and $0.50\pi$ are very close
but the curve of $\theta = 0.48\pi$ is shifted downward slightly when compared with the latter.
For $\theta = 0.50\pi$ it is known that $\mathcal{O}_K^{x}$ vanishes at $g = 1$ when $L\to\infty$ \cite{FengZX2007}.
So it is reasonable for us to believe that it will also go to zero ultimately for $\theta = 0.48\pi$.
As a result, $g = 1$ is still inferred as the critical point for the $A_x$--$A_y$ transition.

\begin{figure}[!ht]
\centering
\includegraphics[width=0.95\columnwidth, clip]{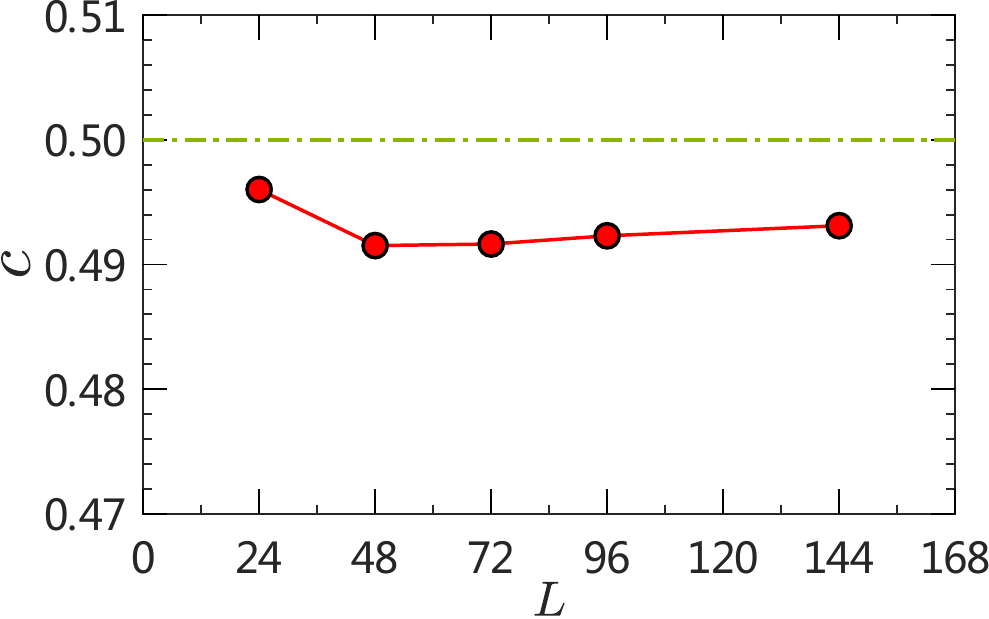}\\
\caption{Estimate of the central charge $c$ for the critical $K$-$\Gamma$ chain at ($g = 1.00$, $\theta = 0.48\pi$).}\label{FIG-Tht048CC}
\end{figure}

To confirm the criticality at $\theta = 0.48\pi$, we now turn to calculate the central charge.
The central charge is usually extracted from the coefficient of the logarithmic correlation
in the entanglement entropy \cite{VidalLRK2003}.
However, this method is not optimal for the critical Kitaev phase
because of the macroscopic ground-state degeneracy \cite{BrzezickiDO2007}.
As a result, it is challenging to get a minimally entangled state
which is essential for a reliable estimate of the central charge.
The practical way to handle this problem is by the energy scaling as shown in Eq.~\eqref{EQ:EgFSS}.
For the PBC, the central charge is given by the following formula
\begin{equation}\label{EQ:CCEgFit}
c_L \simeq \frac{6}{\pi}\big(Le_g-E_g(L)\big)L
\end{equation}
where $e_g$ is the only relevant parameter.
Following a similar procedure illustrated in Fig.~\ref{FIG-GamEg}(b),
we find $e_g\approx-0.1591092$ for $\theta = 0.48\pi$,
which is only slightly larger than that of $-1/(2\pi) = -0.1591549\cdots$ for $\theta = \pi/2$.
We have also calculated the ground-state energy $E_g(L)$ for a series of chain length $L$ ranging from 24 to 144.
The fitting central charge via Eq.~\eqref{EQ:CCEgFit} is shown in Fig.~\ref{FIG-Tht048CC}.
It can be found that the central charge is very close to $1/2$ and suffers from a tiny finite-size effect.
Therefore, we draw the conclusion that the central charge $c = 1/2$,
and the transition belongs to the same universality class as that at $\theta = \pi/2$ \cite{WangCho2015,YangPH2020},
confirming the existence of an extended region of $A_x$ and $A_y$ phases and the critical transition line between them.

\section{The symmetry breaking phases}\label{SEC:SSB}

\subsection{Degeneracy and spin patterns}
Like the AFM case shown in Sec.~\ref{SEC:EHOH},
Eq.~\eqref{J1J2KG-Ham} and Eq.~\eqref{U1XYZ-Ham} could be reduced to
the bond-alternating FM Heisenberg chain when $\theta = \pi/4$ (i.e., $K = \Gamma$).
For this model its ground-state energy $E_g = -(1+g)KL/8$ with a $(L+1)$-fold ground-state degeneracy \cite{KomaNach1997}.
Although it is gapless inherently, the system is not conformally invariant.
Specially, when $g = 1$ it is shown that there is an effective central charge $c_{\rm eff} = 3/2$ \cite{ChenXMetal2013}.
Around the isotropic $SU(2)$ FM point by tuning $\theta$ along the line of $g = 1$, there is an $O_h\to D_4$ symmetry-breaking phase
which has six-fold degenerate ground states along the $\pm\hat{x}$, $\pm\hat{y}$, and $\pm\hat{z}$ spin directions \cite{YangKG2020}.
In addition, the local magnetization, say $\langle \tilde{S}_i^z\rangle$,
shows a three-site periodicity where two of them are equal,
\begin{equation}\label{EQ:FMG1.00}
\langle \tilde{\mathbf{S}}_1\rangle = {c}\hat{z},\; \langle \tilde{\mathbf{S}}_2\rangle = {a}\hat{z},\; \langle \tilde{\mathbf{S}}_3\rangle = {a}\hat{z},
\end{equation}
in which ${a}$ and ${c}$ are the strengths of the local magnetization.
With the $U_6$ transformation shown in Eq.~\eqref{EQ:SOCD} in mind,
it is easy to check that spins in the original basis are
($|S_1^z|$, $|S_2^y|$, $|S_3^y|$; $|S_4^z|$, $|S_5^x|$, $|S_6^x|$) = $(c, a, a; c, a, a)$.
The inherent frustration in model \eqref{U1XYZ-Ham} is accidentally eliminated when $\theta = \pi/4$.
Away from this line, the interplay of bond anisotropy and competing interactions would enhance quantum fluctuations,
giving rise to new type of magnetic orderings.
It is shown in Fig.~\ref{FIG-GSPD} that there are three distinct magnetically ordered states in the middle area
where one is a collinear FM$_{U_6}$ phase while the other two are dubbed $M_1$ and $M_2$ phases.

\begin{figure}[!ht]
\centering
\includegraphics[width=0.95\columnwidth, clip]{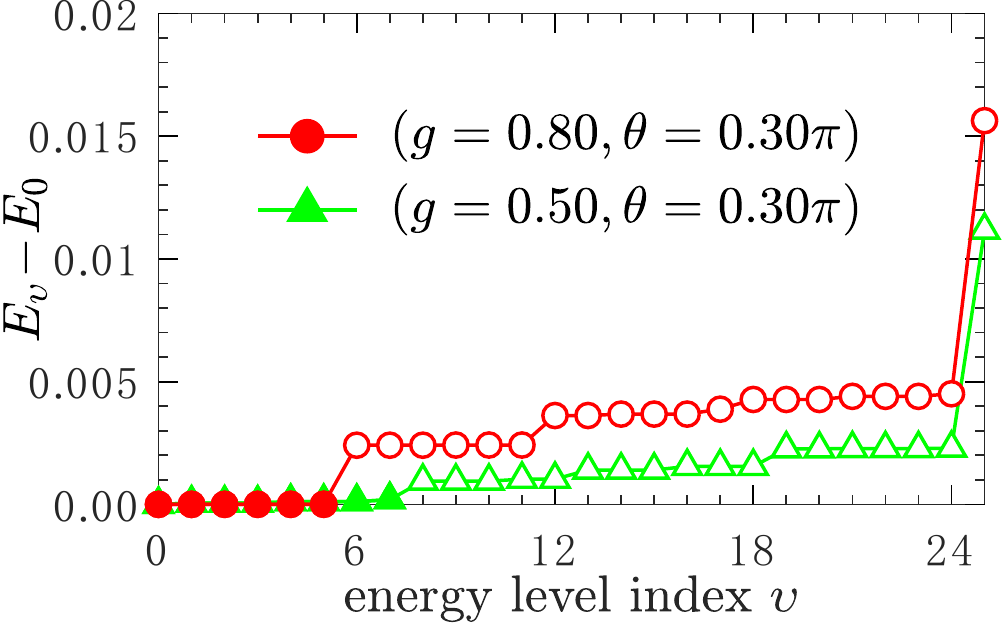}\\
\caption{Low-lying energy levels $E_{\upsilon}$ for $\theta = 0.30\pi$ with chain length $L = 24$.
    The ground-state degeneracy is sixlet and octuplet for $g$ = 0.80 (red circle, FM$_{U_6}$ phase) and $g$ = 0.50 (green triangle, $M_1$ phase), respectively.}
    \label{FIG-FMDegeneracy}
\end{figure}

To begin with, by reducing the strength of $g$ along the line of $\theta = 0.30\pi$,
we find that FM$_{U_6}$ phase could survive until $g\simeq\sqrt3/3$
where the ground-state degeneracy changes from sixlet to octuplet.
Figure~\ref{FIG-FMDegeneracy} show the first ($L$+2) energy levels $E_{\upsilon}$ ($\upsilon$ = $0\to25$) of a 24-site chain
at $g$ = 0.80 (red circle) and 0.50 (green triangle).
One can readily recognize that there is a energy step at the sixth (eighth) energy level for $g$ = 0.80~(0.50).
The energy barrier is $\sim\!10^{-3}$, which is several orders larger than the energy splitting within the degenerate ground states.
The energy step at the 24-th energy level is extremely steep,
which is a reminiscence of the $(L+1)$-fold degeneracy at the $SU(2)$ FM line of $\theta = \pi/4$.
We have checked the ground-state degeneracies under spin chains of $L = 48$ and 72 as well,
and the results remain unchanged in the DMRG calculation with up to $m = 4000$ states kept.

\begin{figure}[!ht]
\centering
\includegraphics[width=0.95\columnwidth, clip]{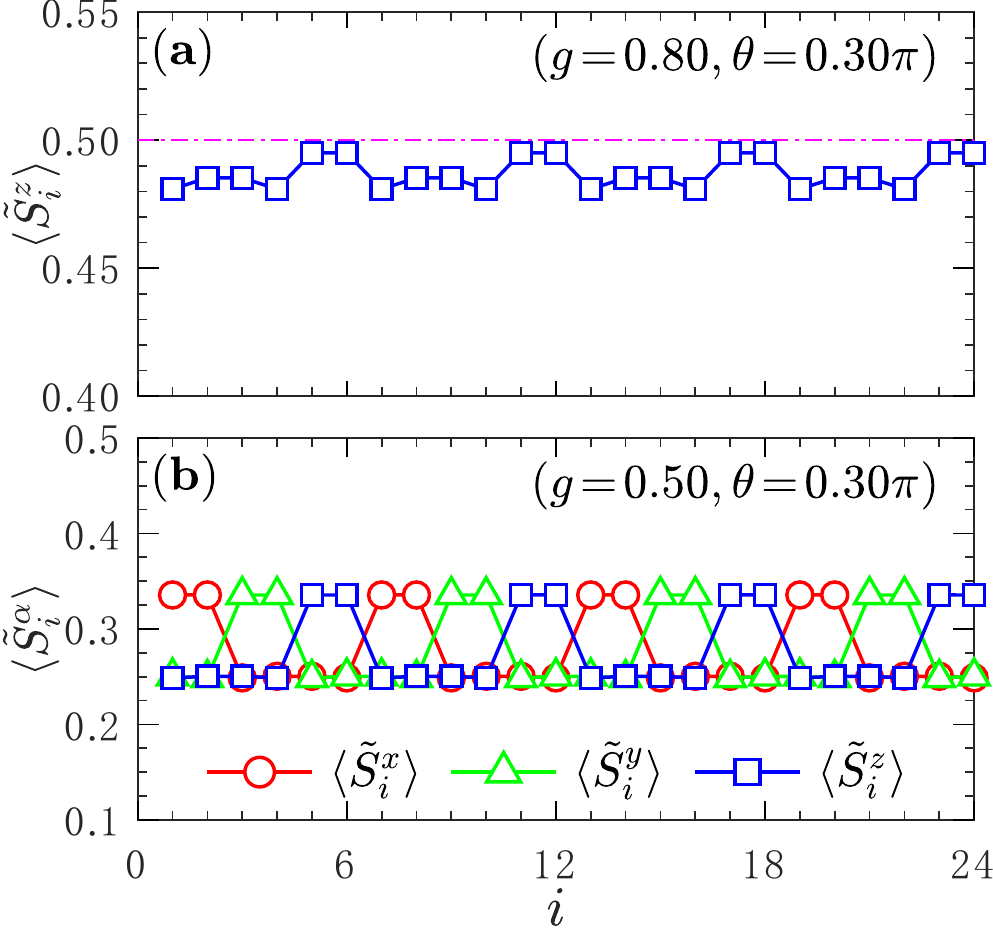}\\
\caption{Local magnetization $\tilde{S}_i^{\alpha}$ ($\alpha$ = $x, y, z$) as a function of site index $i$ for a 24-site chain.
    (a) $\tilde{S}_i^z$ in the FM$_{U_6}$ phase at ($g = 0.80$, $\theta = 0.30\pi$).
    (b) $\tilde{S}_i^x$ (red circle), $\tilde{S}_i^y$ (green triangle), and $\tilde{S}_i^z$~(blue square) in the $M_1$ phase at ($g = 0.50$, $\theta = 0.30\pi$).}
    \label{FIG-FMPattern}
\end{figure}

We then study magnetization distributions of the symmetry breaking phases.
For the FM$_{U_6}$ phase, the spin ordering is very similar to what is shown in Eq.~\eqref{EQ:FMG1.00}
but with a six-site periodicity, see Fig.~\ref{FIG-FMPattern}(a).
It is observed that
\begin{equation}\label{EQ:FMG0.80}
\big(\langle \tilde{\mathbf{S}}_1\rangle, \langle \tilde{\mathbf{S}}_2\rangle, \langle \tilde{\mathbf{S}}_3\rangle; \langle \tilde{\mathbf{S}}_4\rangle, \langle \tilde{\mathbf{S}}_5\rangle, \langle \tilde{\mathbf{S}}_6\rangle\big) \!=\! \big({c},{b},{b};{c},{a},{a}\big)\hat{z}
\end{equation}
where ${a}$, ${b}$ and ${c}$ are magnitudes of the spin orderings along the $\hat{z}$ direction and $a, b, c \leq S$.
There is a slight difference between ${a}$ and ${b}$ when $g \neq 1$.
That is, ${b} < {a}$~(${b} > {a}$) when $g < 1$~($g > 1$).
They are equal at the isotropic case, consistent with the group-theoretical argument \cite{YangKG2020}.
As can be seen from Fig.~\ref{FIG-FMPattern}(a), ${c}$ is the smallest value of the three albeit
its difference to the penultimate value (it is ${b}$ when $g < 1$) becomes negligible as $g$ is decreased.
For the $M_1$ phase shown in Fig.~\ref{FIG-FMPattern}(b),
the $z$ component of the magnetization still shows the pattern in Eq.~\eqref{EQ:FMG0.80},
except that $c$ and $b$ are very close in value but are visibly smaller than $a$.
Most importantly, the $x$ and $y$ components of the spins in the $M_1$ phase also become nonzero
and shows the permutation relation within each even and odd sublattice.
Following the $\eta$-notation of Rousochatzakis and Perkins \cite{RousochatzakisPerkins2017}, we find that
\begin{eqnarray}\label{EQ:N1Spin135}
\langle \tilde{\mathbf{S}}_1\rangle \!=\!
    \left(
    \begin{array}{c}
        \eta_x a \\
        \eta_y b \\
        \eta_z c
    \end{array}
    \right),
\langle \tilde{\mathbf{S}}_3\rangle \!=\!
    \left(
    \begin{array}{c}
        \eta_x c \\
        \eta_y a \\
        \eta_z b
    \end{array}
    \right),
\langle \tilde{\mathbf{S}}_5\rangle \!=\!
    \left(
    \begin{array}{c}
        \eta_x b \\
        \eta_y c \\
        \eta_z a
    \end{array}
    \right)
\end{eqnarray}
and
\begin{eqnarray}\label{EQ:N1Spin246}
\langle \tilde{\mathbf{S}}_2\rangle \!=\!
    \left(
    \begin{array}{c}
        \eta_x a \\
        \eta_y c \\
        \eta_z b
    \end{array}
    \right),
\langle \tilde{\mathbf{S}}_4\rangle \!=\!
    \left(
    \begin{array}{c}
        \eta_x b \\
        \eta_y a \\
        \eta_z c
    \end{array}
    \right),
\langle \tilde{\mathbf{S}}_6\rangle \!=\!
    \left(
    \begin{array}{c}
        \eta_x c \\
        \eta_y b \\
        \eta_z a
    \end{array}
    \right).
\end{eqnarray}
Here, $a, b, c$ $( \geq 0)$ are the intensities of the magnetization,
while $\eta_{x}, \eta_y, \eta_z$ $( = \pm 1)$ are the Ising variables.
It is worth noting that $a$, $b$, and $c$ in the $M_1$ phase satisfy the restriction $\sqrt{a^2+b^2+c^2} \leq S$,
and it is quite different from these in the FM$_{U_6}$ phase (see Eq.~\eqref{EQ:FMG0.80}).
All the three $\eta$'s are free to choose any of 1 or $-1$,
accounting for the eight-fold degeneracy of the $M_1$ phase shown in Fig.~\ref{FIG-FMDegeneracy}.
In addition, by applying the inversion $U_6$ transformation,
the spins in the original basis have the following relation
$(|S_{\upsilon}^x|, |S_{\upsilon}^y|, |S_{\upsilon}^z|) = (a, b, c)$ for $1 \leq \upsilon \leq L$.
The spin structure of the $M_1$ phase is noncoplanar in the rotated basis
and it could be verified by the scalar spin chirality defined as
\begin{equation}\label{EQ:ChiDef}
\hat{\chi}_{ijk} = \tilde{\mathbf{S}}_i \cdot (\tilde{\mathbf{S}}_j \times \tilde{\mathbf{S}}_k).
\end{equation}
It is easy to check that $\hat{\chi}_{135} = \hat{\chi}_{246} \equiv \hat{\chi}_{0}$ and
\begin{align}\label{EQ:ChiNonzero}
\hat{\chi}_{0} &= \eta_x\eta_y\eta_z \big(a^3 + b^3 + c^3 - 3abc\big) \nonumber\\
&= \frac{\eta}{2} (a + b + c)\big[(a - b)^2 + (b - c)^2 + (c - a)^2\big]
\end{align}
with $\eta \equiv \eta_x\eta_y\eta_z$.
Eq.~\eqref{EQ:ChiNonzero} suggests that as long as $a$, $b$, and $c$ are not all the same,
which always holds as observed from Fig.~\ref{FIG-FMPattern}(b),
$\hat{\chi}_{0}$ will be nonzero, in line with the noncoplanar pattern of the $M_1$ phase.

\begin{figure}[!ht]
\centering
\includegraphics[width=0.95\columnwidth, clip]{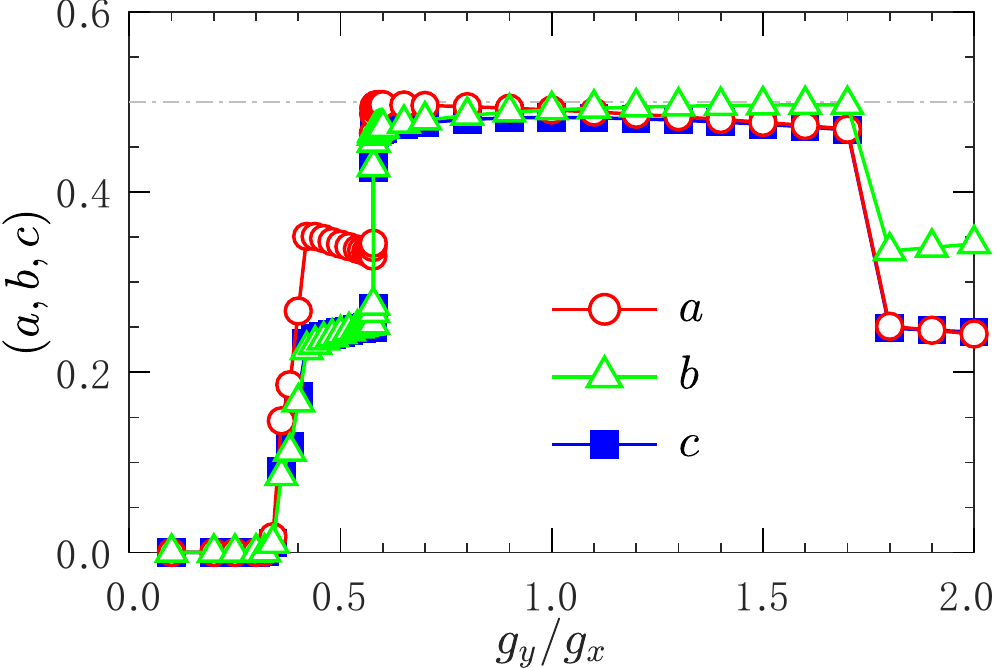}\\
\caption{The $a$ (red circle), $b$ (green triangle), and $c$ (blue square) magnetization components of the FM$_{U_6}$ and $M_1$ phases
    for $\theta = 0.30\pi$ with chain length $L = 48$.
    The middle region is the FM$_{U_6}$ phase ($0.5778 \lesssim g \lesssim 1.731$)
    while the side ones are the self-dual $M_1$ phases.}\label{FIG-FMTht030}
\end{figure}

Figure~\ref{FIG-FMTht030} shows the $(a, b, c)$ components of the magnetization along the line of $\theta = 0.30\pi$.
In the wide region of $0.5778 \lesssim g \lesssim 1.731$,
the ground state is the FM$_{U_6}$ phase where all the spins point along the $z$ direction with a almost saturated moment.
The three species $a$, $b$, and $c$ are totally different as long as $g \neq 1$.
The $M_1$ phase takes over when $0.43 < g < 0.5778$,
and magnitudes of the magnetization are suppressed approximately to $3/4$ (for $a$) or 1/2 (for $b, c$) of the saturated value.

\begin{figure}[!ht]
\centering
\includegraphics[width=0.95\columnwidth, clip]{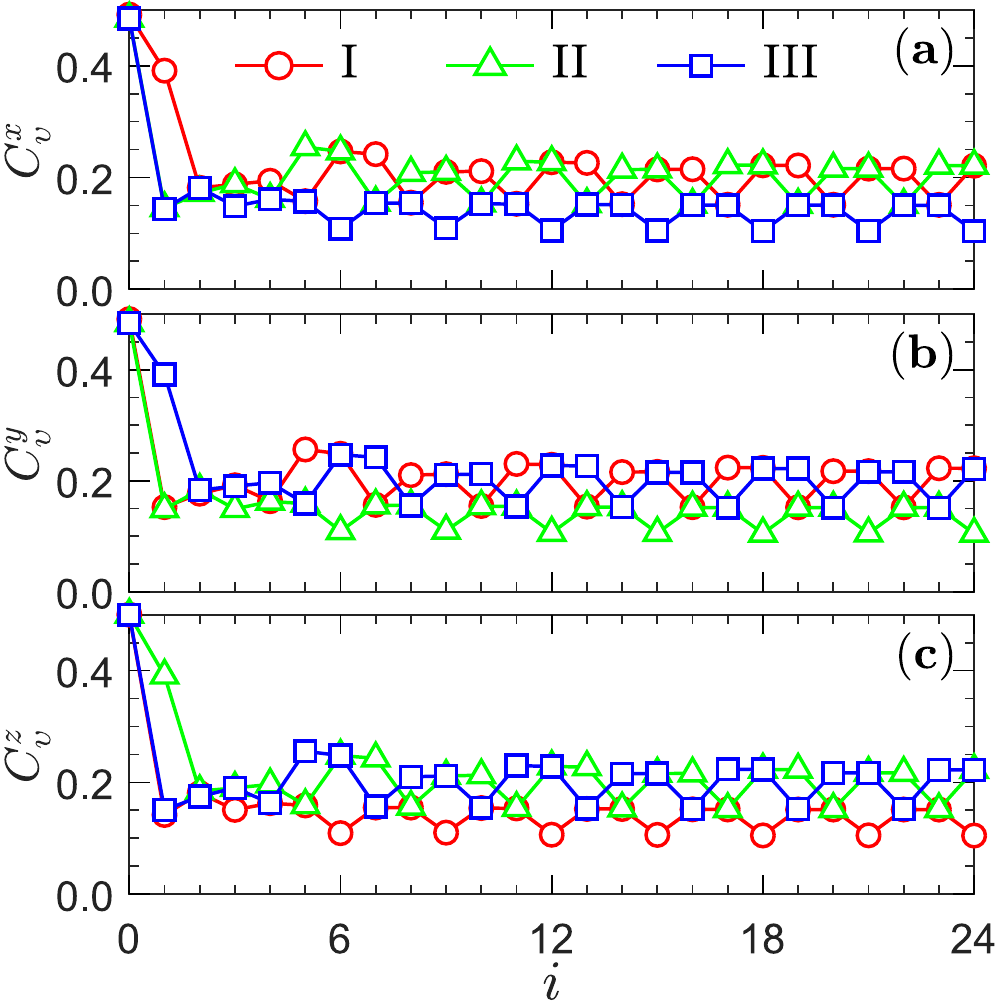}\\
\caption{Spin correlators $C_{\upsilon}^{\alpha}$ ($\alpha$ = $x, y, z$) of a segment $l$ for a 48-site chain.
    The reference site $\upsilon$ could be I (site 1), II (site 2), III (site 3).
    (a), (b), and (c) are for $C^{x}$, $C^{y}$, and $C^{z}$ in the $M_2$ phase at ($g = 1.00$, $\theta = 0.42\pi$), respectively.}
    \label{FIG-CxCyCz}
\end{figure}

For the $M_2$ phase, the local magnetization is fragile and we thus extract their values by calculating the spin-spin correlation functions defined as
\begin{equation}\label{EQ:CorrFunc}
C_{\upsilon}^{\alpha}(l) = \langle \tilde{S}_{\upsilon}^{\alpha} \tilde{S}_{\upsilon+l}^{\alpha} \rangle
\end{equation}
where $\alpha = x, y, z$ and $\upsilon$ is the reference site.
For simplicity we firstly consider the isotropic case ($g$ = 1) which shows a three-site periodicity in the rotated basis.
The correlators $C^{x/y/z}$ at $(g = 1.00, \theta = 0.42\pi)$ are calculated based on a 48-site chain, see Fig.~\ref{FIG-CxCyCz}.
These values are very stable when the site distance $l$ is larger than 10,
and we estimate the local magnetization as $\langle \tilde{S}_{\upsilon}^{\alpha} \rangle = \sqrt{C_{\upsilon}^{\alpha}(L/2)}$ with $L = 48$.
The local magnetization $\langle \tilde{\mathbf{S}}_{\upsilon} \rangle = (\langle \tilde{S}_{\upsilon}^x\rangle, \langle \tilde{S}_{\upsilon}^y\rangle, \langle \tilde{S}_{\upsilon}^z\rangle)^T$
within the 3-site unit cell is
\begin{eqnarray*}
\left(\langle \tilde{\mathbf{S}}_1 \rangle,\langle \tilde{\mathbf{S}}_2 \rangle,\langle \tilde{\mathbf{S}}_3 \rangle\right) \!=\!
    \left(
    \left[\!
    \begin{array}{c}
        0.214 \\
        0.214 \\
        0.103
    \end{array}
    \!\right]\!,
    \left[\!
    \begin{array}{c}
        0.214 \\
        0.103 \\
        0.214
    \end{array}
    \!\right]\!,
    \left[\!
    \begin{array}{c}
        0.103 \\
        0.214 \\
        0.214
    \end{array}
    \!\right]
    \right)\!.
\end{eqnarray*}
When away from the isotropic line where $g = 1$, the unit cell is doubled
and there is a same magnetization distribution to that one shown in Eq.~\eqref{EQ:N1Spin135} and Eq.~\eqref{EQ:N1Spin246}.
It should be noted that both $M_1$ and $M_2$ phases are eight-fold degenerate and their difference lies in the relative values among $a$, $b$, and $c$.
For $M_1$ phase we have $c \simeq b < a$ (when $g < 1$) or $c \simeq a < b$ (when $g > 1$),
while for $M_2$ phase $c$ is much smaller than $a, b$.

\subsection{Magnetic orderings of $M_1$ and $M_2$ phases}
\begin{figure}[!ht]
\centering
\includegraphics[width=0.95\columnwidth, clip]{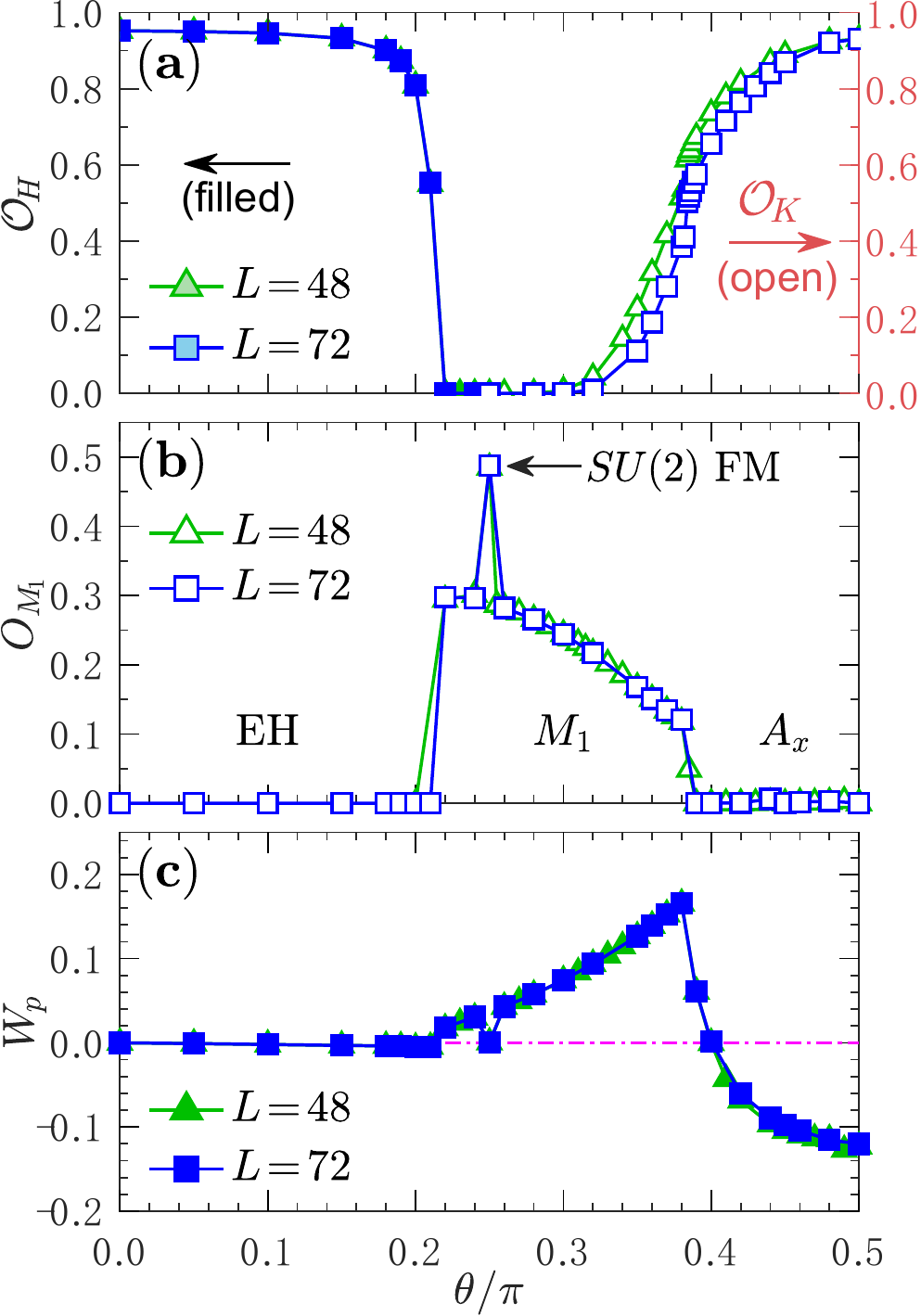}\\
\caption{(a) SOPs of the Haldane-type $\mathcal{O}_H$~(filled symbols) and Kitaev-type $\mathcal{O}_K$~(open symbols)
    for $g = 0.5$ with chain length $L = 48$~(green triangle) and 72~(blue square).
    (b) Order parameter of the $M_1$ phase $O_{M_1}$~($c$ component only) and (c) Flux density $\langle W_p\rangle$
    in the same region as (a).}
    \label{FIG-OPWpJ2050}
\end{figure}

This section is devoted to study the transitions to $M_1$ and $M_2$ phases.
We begin by considering the transitions to $M_1$ phase along the path of $g = 0.5$.
The SOPs of the Haldane-type $\mathcal{O}_H = \mathcal{O}_e^z$ (via Eq.~\eqref{U1XYZ-Ham}) and
the Kitaev-type $\mathcal{O}_K = \mathcal{O}_K^{x}$ (via Eq.~\eqref{KtvGam-Ham}) are plotted in Fig.~\ref{FIG-OPWpJ2050}(a).
At $\theta_H \approx 0.215\pi$ and $\theta_K \approx 0.383\pi$, the two SOPs are discontinuous,
indicating of first-order transitions between the EH~($A_x$) phase and the intermediate $M_1$ phase.
The order parameter $O_{M_1}$ is shown in Fig.~\ref{FIG-OPWpJ2050}(b).
Here, only $c$ flavor of $\langle S_i^z\rangle$ is chosen for the sake of brevity.
We find that it is very robust with a negligible finite-size effect.
In addition, there is also a nonvanishing correlation of a rank-2 spin-nematic (SN) ordering defined in Eq.~\eqref{EQ:SNOrder} (not shown).
Notably, when crossing the line of $\theta = \pi/4$ where the ground state is the $SU(2)$ FM phase,
$O_{M_1}$ has a discontinuity because of the inherent difference of the spin orientations.

Meanwhile, it is appealing to know how the flux density $\langle\hat{W}_p\rangle$,
\begin{equation}\label{EQ:Wp}
\hat{W}_p = 2^6 \tilde{S}_1^z \tilde{S}_2^y \tilde{S}_3^x \tilde{S}_4^z \tilde{S}_5^y \tilde{S}_6^x,
\end{equation}
evolves in each different phase. Similar to the two-dimensional counterpart \cite{Kitaev2006},
the quantity in Eq.~\eqref{EQ:Wp} is the product of spin operators on consecutive overhanging bonds
within the six-site unit cell~(see Fig.~\ref{FIG-Bond}(b)).
In Fig.~\ref{FIG-OPWpJ2050}(c), we plot the flux density $\langle\hat{W}_p\rangle$ versus $\theta$
in the whole region of $\theta \in [0, \pi/2]$.
It is clearly shown that $\langle\hat{W}_p\rangle$ is zero in the EH phase.
In the $M_1$ phase, $\langle\hat{W}_p\rangle$ starts from a nonzero value and goes up with the increasing of $\theta$
except for $\theta = \pi/4$ where $\langle\hat{W}_p\rangle \simeq 0$.
In the $A_x$ phase, however, $\langle\hat{W}_p\rangle$ decreases from 0.17 or so
and does not stop dropping until $\theta \simeq 0.50$ where $\langle\hat{W}_p\rangle < 0$.
Without doubt, the flux density $\langle\hat{W}_p\rangle$ shows a crucial difference among the three distinct phases.
The jump and kink are excellent probes for phase transitions involved.

We now turn to the transition around the $M_2$ phase.
It is shown in the isotropic $K$-$\Gamma$ chain that there is an intermediate phase when $0.40\pi \lesssim \theta \lesssim 0.466\pi$ \cite{YangSN2021}.
This phase is now recognized as the $M_2$ phase with a nonzero magnetization,
and it could survive against small anisotropy where $|g-1| \ll 1$, and then gives way to the conventional collinear FM$_{U_6}$ phase.
Although both phases are magnetically ordered,
we appreciate the rank-2 SN ordering as a sensitive probe to capture the phase transition.
It is finite in the $M_2$ phase while vanishes in the FM$_{U_6}$ phase since the latter is collinear.
The SN correlation function in the long-distance limit is known as \cite{YangSN2021}
\begin{equation}\label{EQ:SNOrder}
O_{SN}^2 = \lim_{|j-i|\gg1}\langle \tilde{S}_{3i+1}^{y}\tilde{S}_{3i+2}^{z} \cdot \tilde{S}_{3j+1}^{y}\tilde{S}_{3j+2}^{z} \rangle.
\end{equation}
Since the species of spins in the correlator of Eq.~\eqref{EQ:SNOrder} always come in pair,
the sign of the correlation remains uninfluenced.
In addition, on can infer from Eq.~\eqref{EQ:N1Spin135} and Eq.~\eqref{EQ:N1Spin246}
that there is no difference of the magnetization among the degenerate ground states,
indicating that all of them will produce a same value of the order parameter.
As a result, we do not need to distinguish these states and
the value of the order parameter could be safely obtained via Eq.~\eqref{EQ:SNOrder}.

\begin{figure}[!ht]
\centering
\includegraphics[width=0.95\columnwidth, clip]{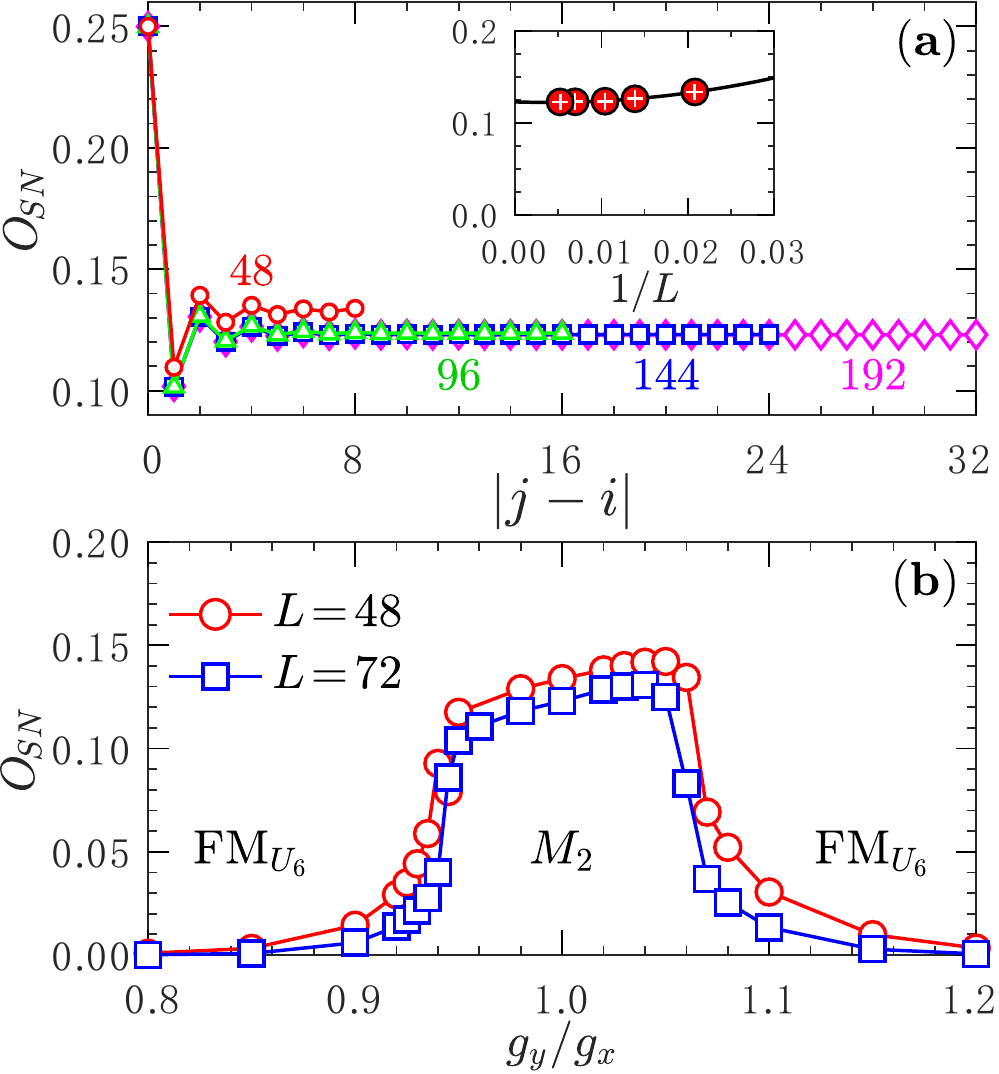}\\
\caption{(a) Correlations of the SN order (see Eq.~\eqref{EQ:SNOrder}) for chain length $L = 48, 96, 144$, and 192.
    The selected point at ($g = 1.00$, $\theta = 0.42\pi$) is deep in the $M_2$ phase.
    Inset: Extrapolation of the SN order to infinite-size system.
    (b) Order parameter $O_{SN}$ of the $M_2$ phase for $\theta = 0.42\pi$
    with chain length $L = 48$~(red circle) and 72~(blue square).}
    \label{FIG-SN}
\end{figure}

Figure~\ref{FIG-SN} shows the order parameter $O_{SN}$ for $\theta = 0.42\pi$.
To check for the finite-size effect, we consider the isotropic case $g = 1$ and
calculate the correlation function in Eq.~\eqref{EQ:SNOrder}
with $i = 0$ and $j$ =0, 1, 2, $\cdots$, $L/6$, see Fig.~~\ref{FIG-SN}(a).
It is found that the correlators saturate to a finite value after several times of oscillation.
In what follows we shall define $O_{SN} \equiv O_{SN}(i=0, j=L/6)$.
The inset shows the extrapolation of the SN order parameter for chain length $L$ = 48, 72, 96, 144 and 192,
from which we can clearly find that $O_{SN}$ is very robust against $L$.
We then extend the calculation of $O_{SN}$ for $0.8 \leq g \leq 1.2$ and the results are summarized in Fig.~~\ref{FIG-SN}(b).
Deep into the $M_2$ phase, $O_{SN}$ is very stable although there is a modest suppression near the boundaries.
The transitions between the $M_2$ phase and the collinear FM$_{U_6}$ phase are accompanied by the jumps of $O_{SN}$,
from which the transition points are determined as $g_t \approx$ 0.945 and 1.065, respectively.
It is worth mentioning that the transition points satisfy the self-dual relation shown in Eq.~\eqref{SelfDual}
since they are related as $ 0.945 \simeq 1/1.065$.

Empirically, the ground-state energy per-site of the $FM_{U_6}$ phase is given by
\begin{equation}\label{EQ:EgFMU6}
e_g^{\textrm{FM}_{U_6}} = -\frac{1}{6}\big[K(a^2+gb^2) + 2c\Gamma(b+ga)\big]
\end{equation}
where $a$, $b$, $c$ are almost saturated (see Fig.~\ref{FIG-FMPattern}(a) and Fig.~\ref{FIG-FMTht030}).
For example, at the hidden $SU(2)$ FM point where $K = \Gamma$ and $g = 1$,
we have $a = b = c = 1/2$ and the energy inferred from Eq.~\eqref{EQ:EgFMU6} is $-K/4$,
consistent with the analytical result \cite{KomaNach1997}.
For the $M_2$ (and also $M_1$) phase, the energy displays a very similar form except that
the prefactor ($1/6$) in Eq.~\eqref{EQ:EgFMU6} should be $1/2$.
In addition, $(a, b, c)$ subject to the constraint $\overline{M} \equiv \sqrt{a^2 + b^2 + c^2} \leq S$.
However, the total magnetization $\overline{M}$ of the $M_2$ phase is far from saturated,
and it is only 0.320 with $(a, b, c) = (0.214, 0.214, 0.103)$ at $(g = 1.00, \theta = 0.42\pi)$.
By adding the bond alternation with $g \neq 1$,
there is a slight enhancement of $\overline{M}$, lowering the ground-state energy and thus opening a finite region of $M_2$ phase.

\section{Transitions around the FM Kitaev limit}\label{SEC:FMKtv}

So far, we have mainly concentrated on the right panel of the phase diagram shown in Fig.~\ref{FIG-GSPD}.
Phases in the left panel could be obtained from the right part after a mirror operation.
However, little is known about the transition types of the adjacent phases near the axis of symmetry.
After an inspection of the first-order energy derivative $\partial e_g/\partial\theta$ along the line of $g = 1$,
Yang \textit{et. al.} claimed that the transition at the FM Kitaev point is of first order \cite{YangKG2020}.
A variational Monte Carlo study, amazingly, suggests that the $Z_2$ QSL at that point
could survive up to a small $\Gamma$-interaction \cite{WangLiu2020}.
Herein, we find that the FM Kitaev point is a confluence point of two transition lines,
i.e., the $A_x$-$A_y$ and the LL-LL$'$ transition lines.
It is thus a multicritical point which accounts for the difficulty in determining the nature of transition
(for extended discussion, see Sec. III in the Supplemental Material \cite{SuppMat}).
By virtue of an efficient bond-reversal method \cite{LuoQPT2019},
we argue in the following that the aforesaid topological QPT between the two LLs is continuous.
Nevertheless, away from the symmetric line of $g = 1$, transitions at $\theta = -\pi/2$ are of first order
without closing the gap at the transition points.

Here we illustrate how to use the bond-reversal method to determine the transition type around the FM Kitaev point.
By tuning $\Gamma$ from negative to positive,
the ground-state energy $e_g$ must be symmetric with respect to the $\Gamma = 0$ line
(i.e., $\theta = -\pi/2$) due to the symmetry relation of Eq.~\eqref{GammaSym}.
When $\theta$ is slightly away from $-\pi/2$,
the sign of $\Gamma$-interaction is changed,
and local expectations of $\Gamma_x = \langle S_1^yS_2^z + S_1^zS_2^y\rangle$
and $\Gamma_y = \langle S_2^zS_3^x + S_2^xS_3^z\rangle$ in the \textit{original} basis must be reversed.
In this regard, we thus define the difference of bond strength~(DBS) $\mathcal{D}$ as
\begin{equation}\label{EQ:DBS}
\mathcal{D} = \frac{1}{2}(\Gamma_x + \Gamma_y).
\end{equation}
The DBS $\mathcal{D}$ is a sensitive probe for a first-order QPT because it have a jump at the transition point.
Oppositely, there is a continuous QPT if $\mathcal{D}$ is smoothly changed \cite{LuoQPT2019}.
We note in passing that, physically, $\mathcal{D}$ is equivalent to
the first-order derivative of ground-state energy $\partial e_g/\partial\theta$ when $g = 1$.
However, the energy derivative depends on the increment $\delta\theta$ which may cause artificial oscillation.
In this sense, the DBS $\mathcal{D}$ is obviously superior and is more reliable.

\begin{figure}[!ht]
\centering
\includegraphics[width=0.95\columnwidth, clip]{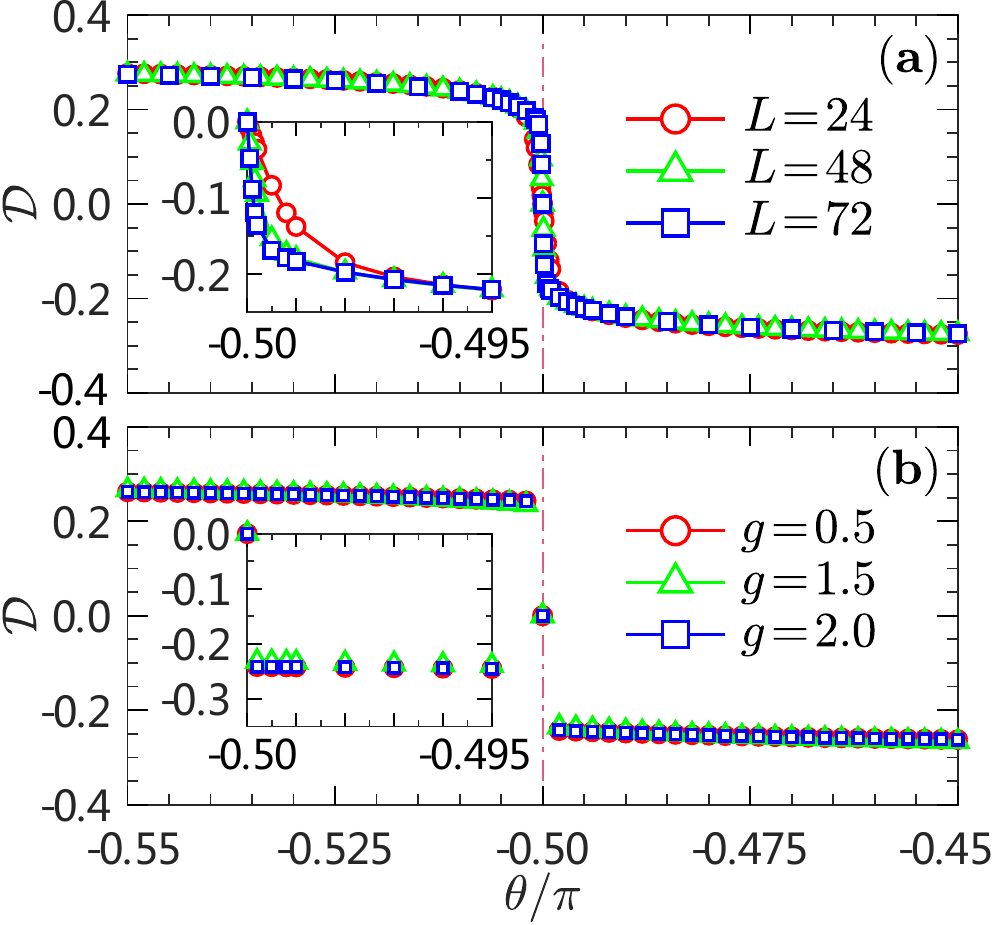}\\
\caption{(a) DBS $\mathcal{D}$ of the $g_x$-$g_y$ $K$-$\Gamma$ chain for $g$ = 1.0 with
    chain length $L = 24$~(red circle), 48~(green triangle) and 72~(blue square).
    Inset: Zoom in of the DBS near the FM Kitaev limit with $\Gamma > 0$.
    (b) DBS $\mathcal{D}$ for $g$ = 0.5~(red circle), 1.5~(green triangle) and 2.0~(blue square) with chain length $L = 24$.}
    \label{FIG-KtvDBS}
\end{figure}

Figure~\ref{FIG-KtvDBS}(a) shows the DBS $\mathcal{D}$ of the LL--LL$'$ transition when $g$ = 1.0.
It is rather smooth without any jump in a wide region of $-0.55\pi \leq \theta \leq -0.45\pi$.
The size-dependent behavior is insignificant except for a narrow slit near the FM Kitaev point.
As can be seen from the inset of Fig.~\ref{FIG-KtvDBS}(a),
the DBS $\mathcal{D}$ shows a well-controlled scaling behavior for different chain length $L$,
and it does not have a jump although its slope becomes sharp as $L$ increased,
indicating of a multicritical behavior.
We recall that such a multicritical point is analogous to the one existing in a 1D transverse $XY$ spin chain
which owns an intersection point of two transition lines of different universality classes
\cite{LiebSM1961,DamleSach1996,YangZS2019}.
When shifting away from $g$ = 1.0, the ground state is occupied by gapped EH or OH phase.
The DBS $\mathcal{D}$ for three selected EH--EH$'$ transition~($g$ = 0.5) and OH--OH$'$ transitions~($g$ = 1.5 and 2.0)
are shown in Fig.~\ref{FIG-KtvDBS}(b).
For all the cases there are appreciable jumps of $\mathcal{D}$ at $\theta = -\pi/2$,
representing the hallmark character of the first-order QPT.

\section{Summary and Discussion}\label{SEC:Conclusion}
We have numerically studied the phases and phase transitions in a bond-alternating spin-$1/2$ $K$-$\Gamma$ chain,
which is an excellent platform to reveal many aspects of one-dimensional quantum magnetism.
By calculating various conventional symmetry-breaking order parameters and nonlocal SOPs,
we unveil a rich quantum phase diagram which contains seven different phases.
Near the AFM Kitaev spin chain limit, there is a critical segment with a macroscopic ground-state degeneracy.
It is unstable against bond alternation, resulting in two gapped disordered $A_x$ and $A_y$ phases characterized by nonlocal SOPs.
The $A_x$--$A_y$ topological QPT falls in the Ising universality class with a central charge $c = 1/2$.
On the other hand, starting from the FM Kitaev spin chain limit by increasing the $\Gamma$ interaction,
there are EH and OH phases in the inner circle ($g < 1$) and outer circle ($g < 1$), respectively.
The EH--OH transition is determined by the SOPs which vanish algebraically at the transition boundary.
It could also be captured by the entanglement gap which undergoes a sign change when crossing the critical point.
This transition belongs to the Gaussian universality class with a central charge $c = 1$,
identical to that of the bond-alternating spin-$1/2$ AFM Heisenberg chain.
The FM Kitaev point is recognized as a multicritical point converging several different phases.
In addition, there are also three distinct magnetically ordered states, named FM$_{U_6}$, $M_1$, and $M_2$ phases,
in the presence of AFM Kitaev interaction.
The FM$_{U_6}$ phase has a six-fold degeneracy and is situated in a wide region around the isotropic line of $g = 1$.
The $M_1$ and $M_2$ phases are highly spatially modulated and could have a rank-2 spin-nematic ordering.

The isotropic $\Gamma$-chain is conformally invariant with a central charge $c = 1$.
While its ground-state energy smoothly varies with the chain length under OBC,
it surprisingly shows an unconventional six-site periodicity under PBC.
We remark that this phenomenon has a profound relation to the abnormal energy scaling in two dimensional honeycomb lattice \cite{LuoZhaoKeeWang2019}.
Given that there is an emergent $SU(2)$ symmetry at this $\Gamma$ limit \cite{YangKG2020},
we conjecture that the versatile Bethe ansatz may be capable to give an exact solution of the isotropic $\Gamma$-chain.

In closing, our work demonstrates the essential role played by the bond alternation in enriching the underlying phase diagram.
The bond alternation is a relevant perturbation to either open up the energy gap or rearrange the distribution of magnetization,
leaving the possibility for the emergence of novel phases.
Our study also highlights the richness of Kitaev systems with AFM exchange interaction.
Although the $K$-$\Gamma$ model is widely recognized as a cornerstone
to describe candidate Kitaev materials like $\alpha$-RuCl$_3$,
much less attention has been paid to $K > 0$ as Kitaev interaction is likely negative in these materials.
A theoretical proposal for the AFM Kitaev interaction in $f$-electron based magnets has been proposed \cite{JangSKM2019}.
Our study thus corroborates a new direction to hunt for exotic phases in a less explored area.

\begin{acknowledgments}
Q.L. would like to thank W.-L. You for fruitful discussions on the quantum compass model.
X.W. was supported by the National Program on Key Research Project (Grant No. 2016YFA0300501)
and the National Natural Science Foundation of China (Grant No. 11974244),
He also acknowledged the support from a Shanghai talent program.
J.Z. was supported by the National Natural Science Foundation of China (Grant No. 11874188).
H.-Y.K. was supported by the NSERC Discovery Grant No. 06089-2016,
the Centre for Quantum Materials at the University of Toronto,
the Canadian Institute for Advanced Research,
and also a funding from the Canada Research Chairs Program.
Computations at early stages were performed on the Tianhe-2JK at the Beijing Computational Science Research Center (CSRC).
Computations were mostly performed on the Niagara supercomputer at the SciNet HPC Consortium.
SciNet is funded by: the Canada Foundation for Innovation under the auspices of Compute Canada;
the Government of Ontario; Ontario Research Fund - Research Excellence; and the University of Toronto.
\end{acknowledgments}

%




\clearpage

\onecolumngrid

\newpage

\newcounter{equationSM}
\newcounter{figureSM}
\newcounter{tableSM}
\stepcounter{equationSM}
\setcounter{section}{0}
\setcounter{equation}{0}
\setcounter{figure}{0}
\setcounter{table}{0}
\setcounter{page}{1}
\makeatletter
\renewcommand{\theequation}{\textsc{sm}-\arabic{equation}}
\renewcommand{\thefigure}{\textsc{sm}-\arabic{figure}}
\renewcommand{\thetable}{\textsc{sm}-\arabic{table}}


\begin{center}
{\large{\bf Supplemental Material for\\ ``Unveiling the phase diagram of a bond-alternating spin-$\frac12$ $K$-$\Gamma$ chain''}}
\end{center}
\begin{center}
Qiang Luo$^{1}$, Jize Zhao$^{2,\,3}$, Xiaoqun Wang$^{4,\,5}$, and Hae-Young Kee$^{1,\,6}$  \\
\quad\\
$^1$\textit{Department of Physics, University of Toronto, Toronto, Ontario M5S 1A7, Canada}\\
$^2$\textit{Lanzhou Center for Theoretical Physics, Lanzhou University, Lanzhou 730000, China}\\
$^3$\textit{School of Physical Science and Technology $\&$ Key Laboratory for Magnetism and\\
    Magnetic Materials of the MoE, Lanzhou University, Lanzhou 730000, China}\\
$^4$\textit{Key Laboratory of Artificial Structures and Quantum Control (Ministry of Education),\\
    School of Physics and Astronomy, Tsung-Dao Lee Institute,\\
    Shanghai Jiao Tong University, Shanghai 200240, China}\\
$^5$\textit{Collaborative Innovation Center for Advanced Microstructures, Nanjing 210093, China}\\
$^6$\textit{Canadian Institute for Advanced Research, Toronto, Ontario, M5G 1Z8, Canada}
\end{center}

\vspace{1.00cm}

\twocolumngrid

This Supplemental Material contains four Sections which aim to support the conclusions in the main text.
In Sec.~\ref{SMSEC:KTV} we briefly review the main process in deriving the dispersion relation of the Kitaev spin chain.
With the help of the analytical energy, we show the central charge (see Tab.~\ref{TabSM-CC})
and the energy density behavior (see Fig.~\ref{FIG-KtvEgScaling}) in the Kitaev spin chain limit.
Next, in Sec.~\ref{SMSEC:BAHC} we present the string order parameters (SOPs) and energy gap of the bond-alternating spin-$1/2$ Heisenberg chain (see Fig.~\ref{FIGSM-HsbgSOPGap}).
Sec.~\ref{SMSEC:MultCP} is an extended discussion of the multicritical point in the ferromagnetic Kitaev limit.
Finally, for the sake of motivated readers, we give a short summary of the finding in the main text in Sec.~\ref{SMSEC:MainRslt}.

\section{Critical Kitaev spin chain: central charge and energy density behavior}\label{SMSEC:KTV}

In the absence of the $\Gamma$-interaction,
the model in the main text is reduced to the well-known Kitaev spin chain
(a.k.a. the one-dimensional quantum compass model \cite{SM:BrzezickiDO2007,SM:YouTian2008}).
Its Hamiltonian reads
\begin{align}\label{EQ:QCMHz-Ham}
\mathcal{H} = \sum_{i = 1}^{L/2} \big(g_x \sigma_{2i-1}^{x}\sigma_{2i}^{x} + g_y \sigma_{2i}^{y}\sigma_{2i+1}^{y}\big)
\end{align}
where $\boldsymbol{\sigma}_i = \big(\sigma_i^x, \sigma_i^y, \sigma_i^z\big)$ is the Pauli operator at site $i$,
and $g_x$~($g_y$) is the interacting strength for the odd~(even) bond.
$L( = 2L')$ is the length of the chain and PBC is employed, i.e., $\boldsymbol{\sigma}_{L+1}$ = $\boldsymbol{\sigma}_1$.
The model could be diagonalized by the Jordan-Wigner transformation,
which transforms the spin operators to free fermions.
It is found that
\begin{align}\label{EQ:QCMHz-Diag}
\mathcal{H} = \sum_k 2\varepsilon_k \left(\gamma_k^{\dagger}\gamma_k - \frac12\right)
\end{align}
where $\gamma_k$ is the quasiparticle for each $k$, and
\begin{equation}\label{EQ:QCMHz-Disp}
\varepsilon_k = \sqrt{g_x^2+g_y^2 - 2g_xg_y\cos k}
\end{equation}
is the dispersion relation.
the momentum $k$ depends on the boundary condition of the fermion and for PBC we have
\begin{equation}\label{EQ:QCMHz-MomK}
k = 0, \pm\frac{2\pi}{L'}, \pm2\frac{2\pi}{L'}, \cdots, \pi
\end{equation}
with $L'$ quasiparticle states in total.
Namely, we have $k = 2\pi p/L'$ with $p = -L'/2+1, \cdots, L'/2-1, L'/2$.
With the symmetric relation of Eq.~\eqref{EQ:QCMHz-Disp} in mind,
we have the total ground-state energy
\begin{equation}\label{EQ:QCMHz-EgSUM}
E_g(L) = -\sum_k \varepsilon_k = -\sum_{p=-L'/2+1}^{L'/2} \varepsilon_p = -\sum_{q=1}^{L'} \varepsilon_q
\end{equation}
and the energy per site is
\begin{align}\label{EQ:QCMHz-EgInf}
e_{g}
&= -\lim_{L\to\infty} \frac{1}{L}\sum_k \varepsilon_k = -\frac{1}{2\pi} \int_0^{\pi} dk \varepsilon_k  \nonumber\\
&= -\frac{1}{\pi} (g_x+g_y) E(\tilde{g}),
\end{align}
where $\tilde{g} = \frac{2\sqrt{g_xg_y}}{g_x+g_y}$ is a dimensionless parameter,
and $E(m) = \int_0^{\pi/2} dx \sqrt{1-m^2\sin^2x}$ is the complete elliptic integral of the second kind.

The system locates at a conformally invariant critical point when $g_x = g_y = 1$.
It is known from Eq.~\eqref{EQ:QCMHz-EgInf} that $e_g = -2/\pi$.
If we rewrite the operators in Hamiltonian from Pauli Matrix $\boldsymbol{\sigma}_i$ to spin-1/2 operator $\textbf{S}_i = \boldsymbol{\sigma}_i/2$,
the ground state energy is then $e_g = -1/(2\pi)$.
According to the energy scaling formula under the PBC, it is known that \cite{SM:BloteCN1986,SM:Affleck1986}
\begin{equation}\label{EQ:QCMEgFSS}
E_g(L) = -\frac{1}{2\pi}L - \frac{\pi c}{6L} + \mathcal{O}(L^{-2})
\end{equation}
where $c = 1/2$ is the central charge.
In other words, the central charge is given by the following formula
\begin{equation}\label{EQSM:CCEgFit}
c_L \simeq \frac{6}{\pi}\big(Le_g-E_g(L)\big)L.
\end{equation}
Table~\ref{TabSM-CC} shows the central charge $c_L$ for different length $L$ ranging from 12 to 72.
It is found that $c_L$ suffers a weak finite-size effect and is very close to 1/2.

\begin{table}[th!]
\caption{\label{TabSM-CC}
The central charge $c$ fitted by Eq.~\eqref{EQSM:CCEgFit} at the critical (isotropic) Kitaev spin chain.}
\begin{ruledtabular}
\begin{tabular}{ c c  c | c  c c}
$L$ & $E_g(L)$ & $c_L$ & $L$ & $E_g(L)$ & $c_L$\\
\colrule
12   & -1.931\,851\,653\;  & 0.504027   & 48   & -7.644\,894\,149\;   & 0.500250 \\
24   & -3.830\,648\,788\;  & 0.501001   & 60   & -9.553\,661\,305\;   & 0.500160 \\
36   & -5.736\,856\,623\;  & 0.500445   & 72   & -11.369\,207\,475\;  & 0.500111 \\
\end{tabular}
\end{ruledtabular}
\end{table}

\begin{figure}[!ht]
\centering
\includegraphics[width=0.95\columnwidth, clip]{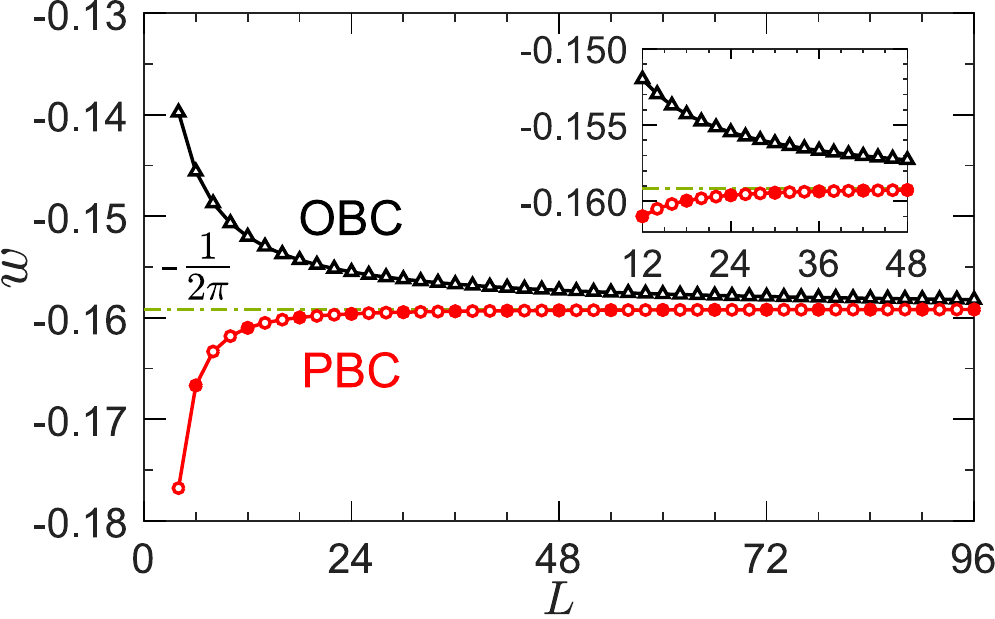}\\
\caption{Behaviors of the energy density $w = E_g(L)/L$ for the isotropic Kitaev spin chain under OBC~(black) and PBC~(red).
    Inset: zoom in of $w$ for $12 \leq L \leq 48$.}
    \label{FIG-KtvEgScaling}
\end{figure}

We also show the energy density of the Kitaev spin chain versus $L$ for OBC and PBC in Fig.~\ref{FIG-KtvEgScaling}.
The energy is obtained by the DMRG calculation and it saturates to $-1/(2\pi)$
(we use spin operator $\textbf{S}_i$ instead of Pauli operator $\boldsymbol{\sigma}_i$).
It is smoothly changed for either OBC or PBC case.
In contrast, Fig.~\textcolor{red}{5}(a) in the main text shows that the energy density $E_g(L)/L$
of the \textbf{critical $\Gamma$-chain} exhibits an oscillating behavior with $L$.
We stress that this is a striking character of the $\Gamma$-chain.

\section{Bond-alternating Heisenberg chain}\label{SMSEC:BAHC}

\begin{figure}[!ht]
\centering
\includegraphics[width=0.95\columnwidth, clip]{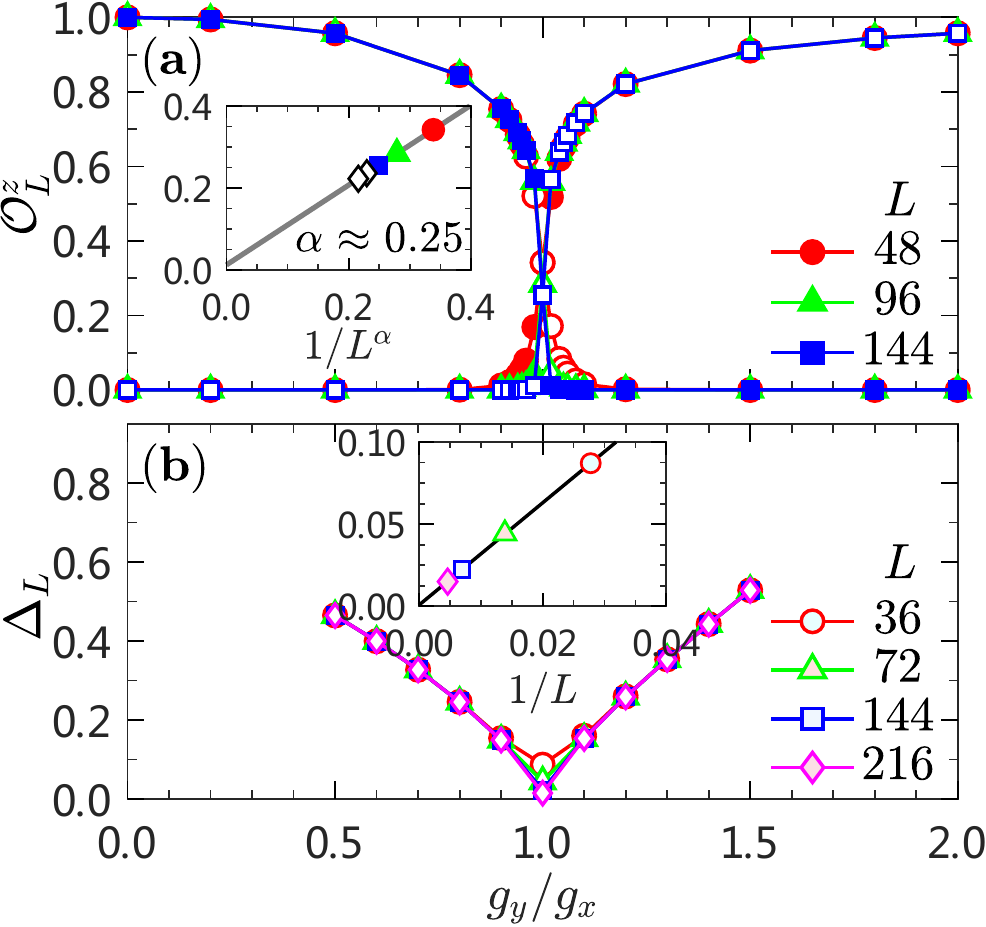}\\
\caption{(a) SOPs of the even type $\mathcal{O}_{e}^z$ (open symbols) and odd type $\mathcal{O}_{o}^z$ (filled symbols)
    for the $g_x$-$g_y$ $K$-$\Gamma$ chain with $\theta= -\pi/4$.
    The inset shows the asymptotic decay of SOP $\mathcal{O}$ at $g_x = g_y$.
    (b) Energy gap $\Delta_L$ in the same region as (a).
    Inset: Linear extrapolation of the energy gap at $g_x = g_y$.}\label{FIGSM-HsbgSOPGap}
\end{figure}

For the bond-alternating $g_x$-$g_y$ $K$-$\Gamma$ chain shown in the main text,
it is equivalent to the bond-alternating Heisenberg chain when $\theta = -\pi/4$ ($|K| = |\Gamma|$),
\begin{equation}\label{EQ:BAHC}
\mathcal{H} = |K|\sum_{i = 1}^{L} \big( g_x \hat{\mathbf{S}}_{2i-1}\cdot \hat{\mathbf{S}}_{2i} + g_y \hat{\mathbf{S}}_{2i}\cdot \hat{\mathbf{S}}_{2i+1} \big).
\end{equation}
For this model, it undergoes a topological QPT between even-Haldane phase ($g_y < g_x$) and odd-Haldane phase ($g_y > g_x$).
The two phases are gapped and possess nonlocal SOPs of different types.
For details see Sec. III in the main text.
Behaviors of the SOPs and energy gap are shown in Fig.~\ref{FIGSM-HsbgSOPGap},
where the SOPs and the energy gap vanish at the critical point where $g_y/g_x = 1$.

\section{FM Kitaev point as a multicritical point Revisited}\label{SMSEC:MultCP}

In the Sec.~VI in the main text, we demonstrate that the FM Kitaev point is a multicritical point where several QPTs meet.
Specially, for the \textit{isotropic} $K$-$\Gamma$ chain with equal bond strength ($g_y/g_x = 1$),
there are two Luttinger Liquids (LLs) in the left and right sides of the FM Kitaev point ($\theta = -\pi/2$).
Regarding the transition type between the two, however, there are two existing works which have different conclusions.
Yang \textit{et. al.} claimed that the transition is of first order \cite{SM:YangKG2020},
while Wang and Liu stated that there is a finite region of $Z_2$ quantum spin liquid ($Z_2$ QSL) between the two LLs \cite{SM:WangLiu2020}.
i) For the first paper, the main argument is that there is a jump for the first-order \textit{total} energy derivative $\partial_{\theta}E_g(L)$
and the jump goes linear with $L$, suggesting a finite jump of the energy derivative in the thermodynamic limit.
ii) For the second paper, the authors used the variational Monte Carlo (VMC) method to compare the ground-state energy of different trial wave functions.
They found that there is a level crossing between the $Z_2$ QSL and LL
and thus concluded that there is an intervening zone between the two LLs.


Here, we investigate the FM Kitaev region closely, because the energy derivative may look sharp
so one needs enough data points to take the energy derivative.
For the VMC result, there could be a competition between $Z_2$ QSL and LL,
yet no solid evidence is presented for a finite region of $Z_2$ QSL but only at the FM Kitaev point.

\subsection{The $XY$ model as an example}
To make it clear, we take an example of the well-known $XY$ model \cite{SM:LiebSM1961},
\begin{equation}\label{XYModel}
\hat{H} = -\sum_{n=1}^{L}\Big((1+\gamma)S_n^xS_{n+1}^x + (1-\gamma)S_n^yS_{n+1}^y + hS_n^z\Big),
\end{equation}
where $S_n^{\alpha}$~$(\alpha = x,y,z)$ is
the $\alpha$ component of the spin operator acting on site $n$,
$\gamma$ is the anisotropy parameter at the $xy$ plane,
and $h$ is the external field along the $z$ direction.
This model owns two kinds of continuous QPT,
one is an Ising transition ($c = 1/2$) occurring at the line of $|h| = 1$
and the other is an anisotropic transition ($c = 1$) with $\gamma = 0$ and $|h| < 1$.
Here we only consider the anisotropic transition.
For a periodic chain with $L$ sites, the total ground-state energy
\begin{equation*}
E_g(L) = -\frac{1}{2}\sum_{n=1}^{L}\sqrt{(h-\cos\frac{2\pi n}{L})^2 + \gamma^2\sin^2\frac{2\pi n}{L}}.
\end{equation*}
According to the Euler-Maclaurin formula, the leading term of the integral is
\begin{align*}
E_g(L) &\simeq -\frac{1}{2}\int_0^{L}\sqrt{(h-\cos\frac{2\pi x}{L})^2 + \gamma^2\sin^2\frac{2\pi x}{L}}dx    \nonumber\\
&= -\frac{L}{2\pi}\int_0^{\pi}\sqrt{(h-\cos t)^2 + \gamma^2\sin^2t}dt,
\end{align*}
which is proportional to $L$.
We note that the integral could be calculated analytically in terms of complete elliptic integrals.
The first-order derivative of $E_g(L)$ w.r.t. $\gamma$ is
\begin{align}\label{EQ:1stEgDer}
\frac{1}{L}\frac{\partial E_g(L)}{\partial\gamma}
&= -\frac{\gamma}{2\pi}\int_0^{\pi} \frac{\sin^2t}{\sqrt{(h-\cos t)^2 + \gamma^2\sin^2t}} dt.
\end{align}
Clearly, the energy derivative in Eq.~\eqref{EQ:1stEgDer} is an odd function of $\gamma$
and it is zero when $\gamma = 0$.
Specifically, when $h = 0$ we have
\begin{align}\label{EQ:XYaniTranEgDer}
\frac{1}{L}\frac{\partial E_g(L)}{\partial\gamma}
&\simeq \frac{\gamma}{\pi(1-\gamma^2)} \left(1+ \ln \left\vert\frac{\gamma}{4} \right\vert \right),
\end{align}
with $|\gamma|\ll1$.
It tends to vanish from either positive value or negative as $\gamma\to0$.
This means that the first-order energy derivative density crosses the transition point smoothly without jump.
We emphasize that this behavior is very similar to the energy derivative of the $K$-$\Gamma$ chain near the FM Kitaev chain,
equivalently, to the difference of bond strength (DBS) shown in Fig.~\textcolor{red}{14} in the main text.

\begin{figure}[!ht]
\centering
\includegraphics[width=0.95\columnwidth, clip]{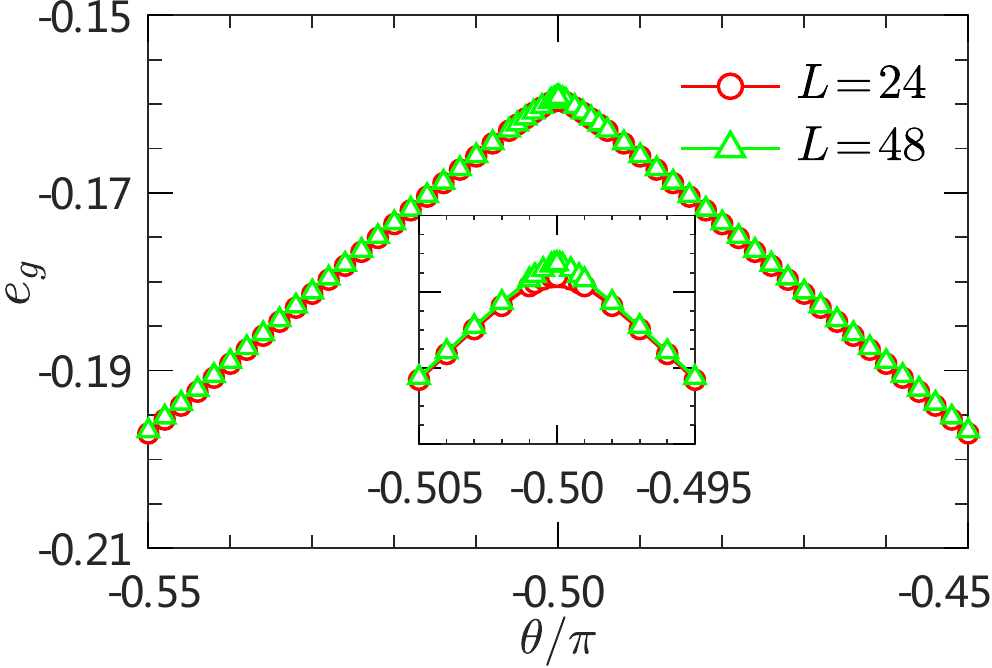}\\
\caption{Ground-state energy $e_g$ for the isotropic $K$-$\Gamma$ chain ($g = 1.00$) near the FM Kitaev point.
    Here, the chain length $L$ = 24 (red circle) and $L$ = 48 (blue square).
    Inset: zoom in of the energy in the very neighboring of the FM Kitaev point.}
    \label{FIGSM-KtvEg}
\end{figure}

\begin{figure}[!ht]
\centering
\includegraphics[width=0.95\columnwidth, clip]{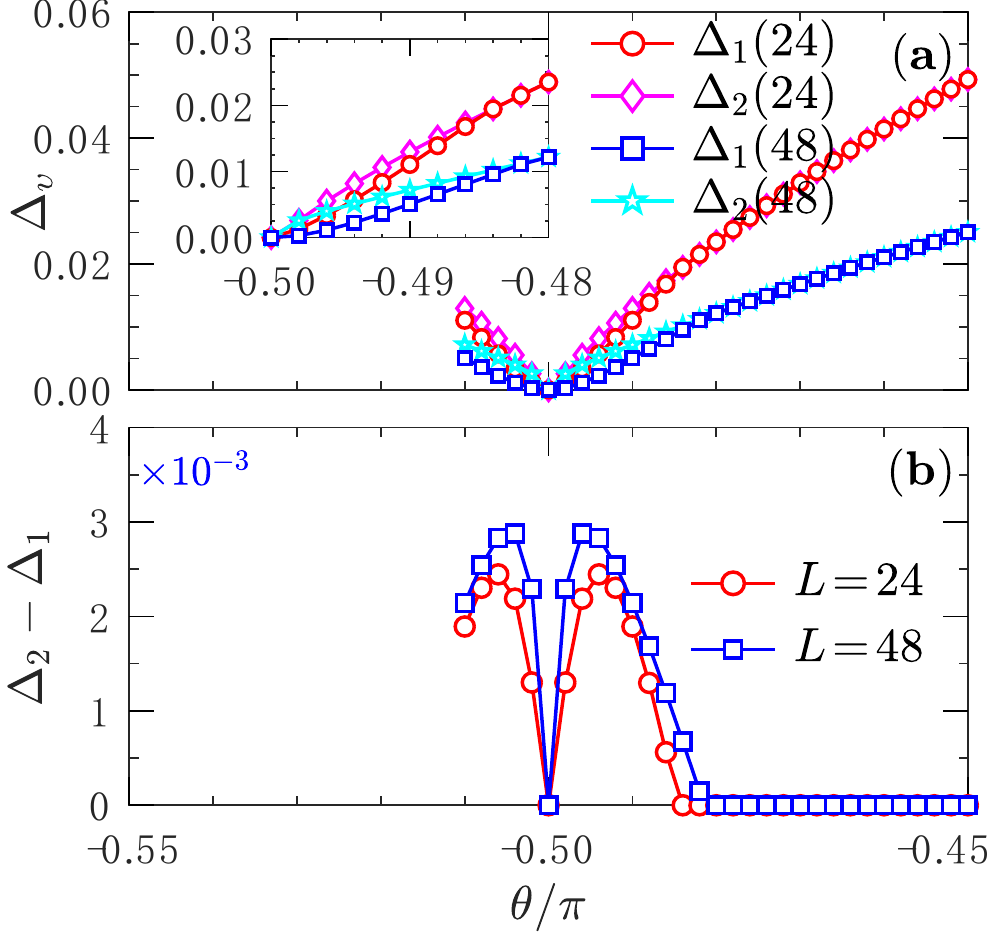}\\
\caption{(a) The first two energy gaps $\Delta_{1,2} = E_{1,2}-E_g$ under the chain length $L$ = 24 (red or pink) and $L$ = 48 (blue or cyan).
    The inset shows the zoom in near the FM Kitaev point.
    (b) Level spacing $\Delta_{2}-\Delta_{1}$ near the FM Kitaev point.}
    \label{FIGSM-KtvGap}
\end{figure}

\subsection{Energy and energy gap near the FM Kitaev point}

As an addendum, we show the ground-state energy $e_g$ in the very vicinity of $\theta = -\pi/2$ (FM Kitaev point)
for the isotropic $K$-$\Gamma$ chain, see Fig.~\ref{FIGSM-KtvEg}.
At the first glance, the energy is symmetric with respect to $\theta = -\pi/2$ (i.e., $E(K, \Gamma) = E(K, -\Gamma)$)
and there is a pinnacle at $\theta = -\pi/2$.
However, if we zoom in the energy in the very surrounding of the FM Kitaev point,
we find that the energy in the top is parabola-like, which is smooth without kink.
Therefore, to get a reliable physics near the multicritical point, one should be careful in interpreting the data.
Numerically, we need to decrease the increment and add enough data points when taking the energy derivative.

In addition, we also calculate the lowest three energy levels along the same line shown in Fig.~\ref{FIGSM-KtvEg}.
The energy gap is defined as $\Delta_{1,2} = E_{1,2}-E_g$,
and the results under the chain length $L$ equals to 24 and 48 are presented in Fig.~\ref{FIGSM-KtvGap}.
The FM Kitaev point has a macroscopic degeneracy and $\Delta_{1,2}$ should be zero at this point.
Upon increasing $\Gamma$-interaction, it is found that the energy gap opens up immediately at the finite-size systems.
This is in line with the consensus that the degeneracy here is very fragile and could be lifted by perturbations.
On the other hand, it also implies that the so-called $Z_2$ QSL does not exist in the presence of even a tiny $\Gamma$-interaction.
As the chain length $L$ increases, the energy gap at $\theta > -\pi/2$ goes down quickly,
giving rise to a gapless LL when $L\to\infty$.
Notably, an interesting observation is that there is an obvious level spacing $\delta\Delta \equiv \Delta_{2}-\Delta_{1} = E_2-E_1$,
between the first and second excited states (see inset).
We also plot the level spacing $\delta\Delta$ in Fig.~\ref{FIGSM-KtvGap}(b),
which shows that $\delta\Delta$ is $\sim\!10^{-3}$ approximately.
The level spacing $\delta\Delta$ is very robust and shows a weak finite-size effect.
The origin and the physical effect of $\delta\Delta$ deserve further studies in future.

\vspace{0.50cm}
\section{Sketch of the main results}\label{SMSEC:MainRslt}

The main findings of the paper is briefly summarized as follows.

\begin{itemize}
  \item \textcolor{blue}{\textsf{I}}. \textbf{One multicritical point}\\
    The isotropic FM Kitaev point at ($\theta = -\pi/2$, $g = 1$) is identified as a multicritical point where the continuous $A_x$-$A_y$ and LL-LL$'$ QPTs meet.
  \item \textcolor{blue}{\textsf{II}}. \textbf{Two topological QPTs}\\
    There are two distinct topological QPTs. One is the Even-Haldane--Odd-Haldane (EH--OH) transition of the Gaussian university class with $c$ = 1,
    The other is the $A_x$--$A_y$ transition of the Ising university class with $c$ = $1/2$.
  \item \textcolor{blue}{\textsf{III}}. \textbf{Three magnetically ordered states}\\
    In the case of AFM Kitaev interaction, there are three magnetically ordered states termed FM$_{U_6}$ phase and $M_1$ and $M_2$ phases.
    For the FM$_{U_6}$ phase, there is a $O_h \to D_4$ symmetry breaking.
    In the $U_6$ rotated basis, the spins are parallel/antiparallel to the $\hat{x}$, $\hat{y}$, or $\hat{z}$ direction,
    and thus the ground-state degeneracy is 2 + 2 + 2 = 6-fold.
    For the $M_1$ and $M_2$ phases, there is a $O_h \to D_3$ symmetry breaking.
    In the $U_6$ rotated basis, the signs of the spins within the unit cell are free to be positive or negative,
    and thus the ground-state degeneracy is $2 \times 2 \times 2$ = 8-fold.
    The difference between the $M_1$ and $M_2$ phases is that the $a$ and $b$ values are relatively different (for definition, see Eqs.~(\textcolor{red}{17}) and (\textcolor{red}{18}) in the main text).
  \item \textcolor{blue}{\textsf{IV}}. \textbf{Four disordered states}\\
    Near the dominating $\Gamma$ region, there are two disordered EH and OH phases which are both gapped.
    They could be characterized by the string order parameters of different kinds.
    The EH phase is topologically trivial while the OH phase is a SPT phase.
    For the latter phase, its ground state is unique (four-fold degenerate) in the case of PBC (OBC).
    In the vicinity of Kitaev limit, there are another two disordered $A_x$ and $A_y$ phases which are also gapped.
    However, the huge ground-state degeneracy is of the order $\mathcal{O}\big(2^{N/2}\big)$ ($N$ is the number of sites).
\end{itemize}


%



\end{document}